\documentclass[12pt]{article}
\pdfoutput=1 

\usepackage[utf8]{inputenc}

\usepackage{amsmath,amsfonts,amssymb,epsfig}
\usepackage{slashed}
\usepackage{cite}
\usepackage{multirow}
\usepackage{cite}

\usepackage{hyperref}

\usepackage{calc}
\usepackage{rotating}
\usepackage[english]{babel}

\usepackage{graphicx}
\usepackage{subfig}
\usepackage{float}
\usepackage{amsmath}
\usepackage{amssymb}
\usepackage{amsthm}
\usepackage{latexsym}
\usepackage{dcolumn}
\usepackage{geometry}
\usepackage{color}
\usepackage{hyperref}
\usepackage{mciteplus}
\usepackage{fancyhdr}

\usepackage[normalem]{ulem}
\usepackage[utf8]{inputenc}
\usepackage{arydshln}
\usepackage{caption}

\hypersetup{
    colorlinks,
    linkcolor={black},
    citecolor={black},
    urlcolor={black}
}

\def\gtwid{\mathrel{\raise.3ex\hbox{$>$\kern-.75em\lower1ex\hbox{$\sim
$}}}}
\def\vio{\mathrel{\hbox{$E$\kern-.60em\hbox{$/
$}}}}
\newcommand{\newc}{\newcommand*}

\long\def\begincomment#1\endcomment{%
        \begingroup\sf\baselineskip12pt#1\endgroup}

\newc{\etal}{\textrm{et al.}} 
\newc{\eg}{\textrm{e.g.}} 
\newc{\ie}{\textrm{i.e.}}
\newc{\etc}{\textrm{etc}}
\newc\vs{\textrm{vs.}}
\newc{\cl}{\rm {C.L.}}

\newc{\ev}{\ensuremath{\,\mathrm{eV}}}
\newc{\kev}{\ensuremath{\,\mathrm{keV}}}
\newc{\mev}{\ensuremath{\,\mathrm{MeV}}}
\newc{\gev}{\ensuremath{\,\mathrm{GeV}}}
\newc{\tev}{\ensuremath{\,\mathrm{TeV}}}
\newc{\MeV}{\mev} 
\newc{\TeV}{\tev}
\newc{\invpb}{\ensuremath{/\text{pb}}}
\newc{\invfb}{\ensuremath{/\text{fb}}}
\newc\nb{\ensuremath{\,\mathrm{nb}}} \newc\pb{\ensuremath{\,\mathrm{pb}}} \newc\fb{\ensuremath{\,\mathrm{fb}}}
\newc\pc{\ensuremath{\,\mathrm{pc}}}
\newc\kpc{\ensuremath{\,\mathrm{kpc}}}
\newc\mpc{\ensuremath{\,\mathrm{Mpc}}}
\newc\ps{\ensuremath{\,\mathrm{ps}}} 
\newc\cmeter{\ensuremath{\,\mathrm{cm}}} 
\newc\meter{\ensuremath{\,\mathrm{m}}} 
\newc\kmeter{\ensuremath{\,\mathrm{km}}}
\newc\second{\ensuremath{\,\mathrm{s}}}
\newc\msecond{\ensuremath{\,\mathrm{ms}}}
\newc\nsecond{\ensuremath{\,\mathrm{ns}}}
\newc\psecond{\ensuremath{\,\mathrm{ps}}}

\newc{\chisqmin}{\ensuremath{\chi^2_{\mathrm{min}}}}
\newc{\Delchisq}{\ensuremath{\Delta\chi^2}}
\newc{\chisq}{\ensuremath{\chi^2}}
\newc{\like}{\ensuremath{\mathcal{L}}}

\newc\lsim{\ensuremath{\mathrel{\rlap{\lower4pt\hbox{\hskip1pt$\sim$}}\raise1pt\hbox{$<$}}}}
\newc\gsim{\ensuremath{\mathrel{\rlap{\lower4pt\hbox{\hskip1pt$\sim$}}\raise1pt\hbox{$>$}}}}
\newc{\VEV}[1]{\ensuremath{\langle #1 \rangle}}
\newc{\dl}{\ensuremath{\stackrel{\leftarrow}{D}}}
\newc{\dr}{\ensuremath{\stackrel{\rightarrow}{D}}}

\newc{\bcenter}{\begin{center}}    \newc{\ecenter}{\end{center}}
\newc{\bfl}{\begin{flushleft}}    \newc{\efl}{\end{flushleft}}
\newc{\bfr}{\begin{flushright}}    \newc{\efr}{\end{flushright}}

\newc{\bi}{\begin{itemize}}
\newc{\ei}{\end{itemize}}
\newc{\bed}{\begin{description}}
\newc{\eed}{\end{description}}
\newc{\ben}{\begin{enumerate}}
\newc{\een}{\end{enumerate}}

\newc{\be}{\begin{equation}}
\newc{\ee}{\end{equation}}
\newc{\bea}{\begin{eqnarray}}
\newc{\eea}{\end{eqnarray}}
\newc{\bfle}{\begin{flalign}}
\newc{\efle}{\end{flalign}}
\newc{\ra}{\rightarrow}

\newc{\alphas}{\ensuremath{\alpha_s}}
\newc{\alphatwo}{\ensuremath{\alpha_2}}
\newc{\alphaone}{\ensuremath{\alpha_1}}
\newc{\alphai}[1]{\ensuremath{\alpha_{#1}}}
\newc{\alphaem}{\ensuremath{\alpha_{\mathrm{em}}}}
\newc{\alphaeff}{\ensuremath{\alpha_{\mathrm{eff}}}}
\newc{\sineff}{\ensuremath{\sin \theta_{\mathrm{eff}}}}
\newc{\sinsqeff}{\ensuremath{\sin^2 \theta_{\mathrm{eff}}}}
\newc{\dalphahad}{\ensuremath{\Delta \alpha_{\mathrm{had}}}}
\newc{\yt}{\ensuremath{h_t}} \newc{\yb}{\ensuremath{h_b}} \newc{\ytau}{\ensuremath{h_{\tau}}}
\newc\mz{\ensuremath{M_Z}} 
\newc\mw{\ensuremath{m_W}}
\newc\mZ{\mz}        \newc\mW{\mw}
\newc\mhsm{\ensuremath{ m_{H_{\mathrm{SM}}}}}
\newc{\mtop}{\ensuremath{ m_t}}               \newc{\mtpole}{\ensuremath{ M_t}}
\newc{\mbottom}{\ensuremath{ m_b}} 
\newc{\mtau}{\ensuremath{ m_{\tau}}}
\newc{\mt}{\mtpole}
\newc{\mb}{\mbottom} 
\newc{\rtwogg}{\ensuremath{R_{h_2}(\gamma\gamma)}}
\newc{\rtwozz}{\ensuremath{R_{h_2}(ZZ)}}
\newc{\ronegg}{\ensuremath{R_{h_1}(\gamma\gamma)}}
\newc{\ronezz}{\ensuremath{R_{h_1}(ZZ)}}
\newc{\rsiggg}{\ensuremath{R_{h_\textrm{sig}}(\gamma\gamma)}}
\newc{\rsigzz}{\ensuremath{R_{h_\textrm{sig}}(ZZ)}}
\newc{\llbar}{\ensuremath{\ell\bar{\ell}}}
\newc{\tauptaum}{\ensuremath{ \tau^+\tau^-}}
\newc{\qqbar}{\ensuremath{ q\bar{q}}} \newc{\ppbar}{\ensuremath{ p\bar{p}}}
\newc{\bbbar}{\ensuremath{ b\bar{b}}} \newc{\ttbar}{\ensuremath{ t\bar{t}}}
\newc{\ffbar}{\ensuremath{ f\bar{f}}} \newc{\tautaubar}{\ensuremath{ \tau\bar{\tau}}}

\newc{\mchi}{\ensuremath{m_\neutone}}
\newc{\squark}{\ensuremath{\tilde{q}}}
\newc{\slepton}{\ensuremath{\tilde{l}}}
\newc{\gluino}{\ensuremath{\tilde{g}}} 
\newc{\mgluino}{\ensuremath{{m_{\gluino}}}}
\newc{\wino}{\ensuremath{\tilde{W}}} 
\newc{\mwino}{\ensuremath{{m_{\wino}}}}
\newc{\tone}{\ensuremath{{\tilde{t}_1}}}
\newc{\Hone}{\ensuremath{{\tilde{H}_{1}}}}
\newc{\Htwo}{\ensuremath{{\tilde{H}_{2}}}}
\newc{\Hhtwo}{\ensuremath{{H_{2}}}}
\newc{\qli}{\ensuremath{{\tilde{Q}_{i}}}}
\newc{\uri}{\ensuremath{{\tilde{u}_{i}}}}
\newc{\dri}{\ensuremath{{\tilde{d}_{i}}}}
\newc{\lli}{\ensuremath{{\tilde{L}_{i}}}}
\newc{\eri}{\ensuremath{{\tilde{e}_{i}}}}

\newc{\sthw}{\ensuremath{ \sin\theta_W}}              \newc{\cthw}{\ensuremath{\cos\theta_W}}
\newc{\tanthw}{\ensuremath{ \tan\theta_W}}              \newc{\cotthw}{\ensuremath{\cot\theta_W}}
\newc{\ssqthw}{\ensuremath{\sin^2 \theta_W}}
\newc{\msbar}{\ensuremath{\overline{MS}}} \newc{\drbar}{\ensuremath{\overline{DR}}}
\newc{\mtmtsmmsbar}{\ensuremath{ m_t(m_t)^{\msbar}_{{\mathrm{SM}}}}}
\newc{\mtmtsmdrbar}{\ensuremath{ m_t(m_t)^{\drbar}_{{\mathrm{SM}}}}}
\newc{\mtmtmssmdrbar}{\ensuremath{ m_t(m_t)^{\drbar}_{{\mathrm{SUSY}}}}}
\newc{\mbmbmsbar}{\ensuremath{ m_b(m_b)^{\msbar} }}
\newc{\mbmbsmmsbar}{\ensuremath{ m_b(m_b)^{\msbar}_{{\mathrm{SM}}}}}
\newc{\mbmzsmmsbar}{\ensuremath{ m_b(\mz)^{\msbar}_{{\mathrm{SM}}}}}
\newc{\mbmzsmdrbar}{\ensuremath{ m_b(\mz)^{\drbar}_{{\mathrm{SM}}}}}
\newc{\mbmzmssmdrbar}{\ensuremath{ m_b(\mz)^{\drbar}_{{\mathrm{SUSY}}}}}
\newc{\mtaumzsmmsbar}{\ensuremath{ m_{\tau}(\mz)^{\msbar}_{{\mathrm{SM}}}}}
\newc{\mtaumzsmdrbar}{\ensuremath{ m_{\tau}(\mz)^{\drbar}_{{\mathrm{SM}}}}}
\newc{\mtaumzmssmdrbar}{\ensuremath{ m_{\tau}(\mz)^{\drbar}_{{\mathrm{SUSY}}}}}
\newc{\alphasmzms}{\ensuremath{\alpha_s(M_Z)^{\overline{MS}}}}
\newc{\alphaimzms}[1]{\ensuremath{\alpha_{#1}(M_Z)^{\overline{MS}}}}
\newc{\alphaemmz}{\ensuremath{\alpha_{\mathrm{em}}(M_Z)^{\overline{MS}}}}

\newc{\mzero}{\ensuremath{{m_0}}}
\newc{\mhalf}{\ensuremath{ m_{1/2}}}
\newc{\tanb}{\ensuremath{\tan\beta}}
\newc{\azero}{\ensuremath{ A_0}}
\newc{\signmu}{\ensuremath{\rm{sgn}\,\mu}}
\newc{\atau}{\ensuremath{{A_{\tau}}}}
\newc{\mueff}{\ensuremath{\mu_{\rm{eff}}}}
\newc{\lam}{\ensuremath{{\lambda}}}
\newc{\kap}{\ensuremath{{\kappa}}}
\newc{\alam}{\ensuremath{{A_{\lambda}}}}
\newc{\akap}{\ensuremath{{A_{\kappa}}}}
\newc{\hs}{\ensuremath{ H_s}}      
\newc{\mhs}{\ensuremath{ m_{H_s}}} 
\newc{\mgut}{\ensuremath{ M_{\rm GUT}}}
\newc{\mvl}{\ensuremath{ M_{\rm VL}}}
\newc{\gut}{\ensuremath{{\rm GUT}}}
\newc{\mplanck}{\ensuremath{ M_{\rm P}}}      \newc{\mpl}{\ensuremath{ M_{\rm Pl}}}
\newc{\msusy}{\ensuremath{ M_{\rm SUSY}}}      \newc{\ms}{\ensuremath{ M_{\rm S}}}
 \newc{\hu}{\ensuremath{ H_u}}       \newc{\hd}{\ensuremath{ H_d}}
 \newc{\mhu}{\ensuremath{ m_{H_u}}}       \newc{\mhd}{\ensuremath{ m_{H_d}}}
 \newc{\mhuew}{\ensuremath{ m^{\ast}_{H_u}}}       \newc{\mhdew}{\ensuremath{ m^{\ast}_{H_d}}}
 \newc{\mhuewsq}{\ensuremath{ m^{\ast\, 2}_{H_u}}}       \newc{\mhdewsq}{\ensuremath{ m^{\ast\, 2}_{H_d}}}
 \newc{\mhl}{\ensuremath{m_\hl}} 
 \newc{\mhone}{\ensuremath{m_{h_1}}} 
 \newc{\mhtwo}{\ensuremath{m_{h_2}}} 
 \newc{\mhi}{\ensuremath{m_{\tilde{h}}}} 
 \newc{\mul}{\ensuremath{m_{\tilde{u}_L}}} 
 \newc{\mtone}{\ensuremath{m_{\tilde{t}_1}}} 
 \newc{\ma}{\ensuremath{m_A}} 
 \newc{\mH}{\ensuremath{m_H}} 
 \newc{\maone}{\ensuremath{m_{a_1}}} 
 \newc{\matwo}{\ensuremath{m_{a_2}}}
 \newc{\hone}{\ensuremath{h_1}}
 \newc{\htwo}{\ensuremath{h_2}}
 \newc{\aone}{\ensuremath{a_1}}
 \newc{\atwo}{\ensuremath{a_2}}
 \newc{\mqthree}{\ensuremath{m_{\tilde{Q}_3}^2}}
 \newc{\muthree}{\ensuremath{m_{\tilde{u}_3}^2}}
 \newc{\mqli}{\ensuremath{m_{\tilde{Q}_{i}}}}
 \newc{\muri}{\ensuremath{m_{\tilde{u}_{i}}}}
 \newc{\mdri}{\ensuremath{m_{\tilde{d}_{i}}}}
 \newc{\mlli}{\ensuremath{m_{\tilde{L}_{i}}}}
 \newc{\meri}{\ensuremath{m_{\tilde{e}_{i}}}}
 \newc{\ts}{\ensuremath{T_{SUSY}}}

\newc{\sigsip}{\ensuremath{\sigma^{\rm SI}_{p}}}	\newc{\sigsin}{\ensuremath{\sigma^{\rm SI}_{n}}}
\newc{\sigsdp}{\ensuremath{\sigma^{\rm SD}_{p}}}	\newc{\sigsdn}{\ensuremath{\sigma^{\rm SD}_{n}}}
\newc{\sigsi}{\ensuremath{\sigma^{\rm SI}}}	\newc{\sigsd}{\ensuremath{\sigma^{\rm SD}}}
\newc{\abund}{\ensuremath{ \Omega h^2}}
\newc{\omegadm}{\ensuremath{ \Omega_{{\rm DM}}}}     \newc{\abunddm}{\ensuremath{ \Omega_{{\rm DM}} h^2}} 
\newc{\omegam}{\ensuremath{ \Omega_{{\rm m}}}}       \newc{\abundm}{\ensuremath{ \Omega_{{\rm m}} h^2}}
\newc{\omegab}{\ensuremath{ \Omega_{{\rm b}}}}	\newc{\abundb}{\ensuremath{ \Omega_{{\rm b}} h^2}}
\newc{\omegatot}{\ensuremath{ \Omega_{{\rm TOT}}}}
\newc{\omegacdm}{\ensuremath{ \Omega_{{\rm CDM}}}}   \newc{\abundcdm}{\ensuremath{ \Omega_{{\rm CDM}} h^2}}
\newc{\omegalambda}{\ensuremath{ \Omega_{\Lambda}}} \newc{\abundlambda}{\ensuremath{ \Omega_{\Lambda} h^2}}
\newc{\omegarad}{\ensuremath{ \Omega_{{\rm rad}}}}  \newc{\abundrad}{\ensuremath{ \Omega_{{\rm rad}} h^2}}
\newc{\rhocrit}{\ensuremath{ \rho_{\rm crit}}}
\newc{\rhochi}{\ensuremath{ \rho_{\chi}}}
\newc{\abunchi}{\ensuremath{\Omega_\chi h^2}}
\newc{\abundlsp}{\ensuremath{\Omega_{\rm LSP}h^2}}

\newc{\amu}{\ensuremath{ a_{\mu}}}        \newc{\amususy}{\ensuremath{ a_{\mu}^{\mathrm{SUSY}}}}
\newc{\amuexpt}{\ensuremath{ a_{\mu}^{\mathrm{expt}}}}        \newc{\amusm}{\ensuremath{ a_{\mu}^{\mathrm{SM}}}}
\newc\deltaamu{\ensuremath{\Delta a_{\mu}}} \newc{\deltaamususy}{\ensuremath{\delta a_{\mu}^{\mathrm{SUSY}}}}
\newc\gmtwo{\ensuremath{ (g-2)_{\mu}}} 
\newc{\deltagmtwomususy}{\ensuremath{\delta\left(g-2\right)_{\mu}^{\mathrm{SUSY}}}}
\newc{\deltagmtwomu}{\ensuremath{\delta\left(g-2\right)_{\mu}}}
\newc{\deltagmtwoe}{\ensuremath{\delta\left(g-2\right)_{e}}}
\newc{\deltagmtwo}{\ensuremath{\delta\left(g-2\right)}}
\newc\BR{\ensuremath{\rm BR}}
\newc\bsgamma{\ensuremath{ b\rightarrow s \gamma }}
\newc\bxsgamma{\ensuremath{\overline{B}\rightarrow X_{s}\gamma}}
\newc\brbsgamma{\ensuremath{\BR\left(\bsgamma\right)}}
\newc\brbxsgamma{\ensuremath{\BR\left(\bxsgamma\right)}}
\newc\bsmumu{\ensuremath{B_s\to\mu^+\mu^-}}
\newc\brbsmumu{\ensuremath{\BR\left(B_s\to\mu^+\mu^-\right)}}
\newc\bdmmumu{\ensuremath{\overline{B}_d\to\mu^+\mu^-}}
\newc\bbbarmix{\ensuremath{\overline{B}_s\mbox{-}B_s}}      
\newc\delmbs{\ensuremath{\Delta M_{B_s}}}
\newc{\butaunu}{\ensuremath{B_u \rightarrow \tau \nu}}
\newc{\brbutaunu}{\ensuremath{\BR\left(B_u \rightarrow \tau \nu\right)}}

\newcommand*{\reftable}[1]{Table~\ref{#1}}         
       
\newcommand*{\reffig}[1]{Fig.~\ref{#1}}
 
        \newcommand*{\refeq}[1]{Eq.~(\ref{#1})}
     \newcommand*{\refsec}[1]{Sec.~\ref{#1}}

\newcommand*{\neutone}{\ensuremath{\tilde{\chi}^0_1}}

\let\oldcite\cite
\renewcommand*{\cite}{~\oldcite}

\newcommand*{\hl}{\ensuremath{h}}

\restylefloat{figure}

\begin{document}

\title{\LARGE {\bf Flavor anomalies \\ from asymptotically safe gravity}}
\author{Kamila Kowalska\footnote{\url{kamila.kowalska@ncbj.gov.pl}},\, Enrico Maria Sessolo\footnote{\url{enrico.sessolo@ncbj.gov.pl}},\, and Yasuhiro Yamamoto\footnote{\url{yasuhiro.yamamoto@ncbj.gov.pl}} \\[2ex]
\small {\em National Centre for Nuclear Research}\\[-2ex]
\small {\em Pasteura 7, 02-093 Warsaw, Poland  }\\
}
%
\date{}
\maketitle
\thispagestyle{fancy}
\begin{abstract}
We use the framework of asymptotically safe quantum gravity to derive predictions for scalar leptoquark solutions to the $b\to s$ and $b\to c$ flavor anomalies.  The presence of an interactive UV fixed point in the system of gauge and Yukawa couplings imposes a set of boundary conditions at the Planck scale, which allows one to determine low-energy values of the leptoquark Yukawa matrix elements. As a consequence, the allowed leptoquark mass range can be significantly narrowed down. We find that a consistent gravity-driven solution to the $b\to s$ anomalies predicts a leptoquark with the mass of $4-7\tev$, entirely within the reach of a future hadron-hadron collider with $\sqrt{s}=100\tev$. Conversely, in the case of the $b\to c$ anomalies 
the asymptotically safe gravity framework predicts a leptoquark mass at the edge of the current LHC bounds.   
Complementary signatures appear in flavor observables, namely the (semi)leptonic decays of $B$ and $D$ mesons and kaons.     
\end{abstract}
\newpage 

\tableofcontents

\setcounter{footnote}{0}

\section{Introduction\label{sec:intro}}

Recent years have seen substantial development in the field of asymptotically safe quantum gravity\cite{inbookWS,Reuter:1996cp,Lauscher:2001ya,Reuter:2001ag,Lauscher:2002sq,Litim:2003vp,Codello:2006in,Machado:2007ea,Codello:2008vh,Benedetti:2009rx,Manrique:2011jc,Dietz:2012ic,Falls:2013bv,Falls:2014tra}.
An ambitious program has emerged around the fact that graviton fluctuations induce in the trans-Planckian regime universal contributions to the renormalization group (RG) running of matter couplings\cite{Robinson:2005fj,Pietrykowski:2006xy,Toms:2007sk,Tang:2008ah,Toms:2008dq,Rodigast:2009zj,Zanusso:2009bs,Daum:2009dn,Daum:2010bc,Folkerts:2011jz,Oda:2015sma,Eichhorn:2016esv,Christiansen:2017gtg,Hamada:2017rvn,Christiansen:2017cxa,Eichhorn:2017eht} 
that can be calculated via the functional renormalization group\cite{WETTERICH199390}. 
It has also been established that, given the matter content of a certain theory under study, at the energy scale where gravity is not negligible the renormalization group system of the gravity and matter couplings can develop a non-trivial (interactive) fixed point\cite{Harst:2011zx,Christiansen:2017gtg,Eichhorn:2017lry,Eichhorn:2017ylw,Eichhorn:2018whv}. This feature has important consequences for our understanding of whether a given quantum field theory can be considered fundamental.  
For example, in the standalone Standard Model (SM) the running of the hypercharge gauge coupling encounters eventually a 
Landau pole in the deep ultraviolet (UV), but the presence of gravitational interactions can tame its running by antiscreening graviton fluctuations, so that the gauge coupling remains finite\cite{Harst:2011zx,Christiansen:2017gtg,Eichhorn:2017lry}.
Thus, the theory with a UV fixed point can be non-perturbatively renormalizable. 
 
The important finding that quantum gravity and matter can feature interactive fixed points in the extreme trans-Planckian regime has opened the exciting possibility of deriving the values of the observable quantities of the SM (gauge, Yukawa and scalar couplings) from first principles\cite{Shaposhnikov:2009pv,Eichhorn:2017ylw,Eichhorn:2018whv,Alkofer:2020vtb}.
The RG flow of the system, emerging from the UV fixed point, down to the electroweak symmetry breaking (EWSB) scale along the UV-safe trajectory, has led in fact to specific predictions (or \textit{post}dictions in some cases) for the top Yukawa coupling\cite{Eichhorn:2017ylw} and the quartic coupling of the Higgs potential\cite{Shaposhnikov:2009pv}. This framework was also used to predict the Lagrangian couplings of a few New Physics (NP) models extending the SM by an extra U(1) gauge symmetry and a scalar field with portal couplings\cite{Grabowski:2018fjj,Kwapisz:2019wrl,Reichert:2019car,Eichhorn:2020kca}.  

In the spirit of the above-cited works, in this paper we try to investigate whether the embedding in an asymptotically-safe gravity framework could improve the predictivity of particular NP models for which some experimental information exists, but it is still incomplete and/or not sufficient to draw a clear direction for future searches that should confirm unequivocally the existence of the NP itself and uncover its properties. 
A practical example we have in mind has to do with the so-called {\it flavor anomalies} in $b\to s$\cite{Aaij:2014ora,Aaij:2014pli,Aaij:2015oid,Aaij:2015esa,Aaij:2016flj,Aaij:2017vbb,Aaij:2019wad,Aaij:2020nrf,Wehle:2016yoi,Abdesselam:2019wac,Sirunyan:2017dhj,Aaboud:2018krd} and $b\to c$\cite{Lees:2012xj,Lees:2013uzd,Huschle:2015rga,Sato:2016svk,Hirose:2016wfn,Aaij:2015yra,Aaij:2017uff,Aaij:2017deq,Abdesselam:2019dgh,Belle:2019rba} transitions. These are a set of measurements, reported over the past several years by LHCb and other experimental collaborations, which suggest to high statistical significance a departure from the SM predictions of certain decay amplitudes involving a change of flavor of the participating fermions. NP solutions to the flavor anomalies are usually cast in terms of bounds on certain products of masses and couplings, but the current experimental information does not allow to pinpoint the particular mass scale at which the NP could be directly observed. An extra piece of theoretical information, provided for example by specific boundary conditions of the NP Yukawa couplings in the deep UV, could help to bridge this gap. 

In this study we focus on the well known scalar leptoquark (LQ) solutions to the flavor anomalies (for a review, see, e.g., Ref.\cite{Dorsner:2016wpm}). Scalar LQs provide a simple, single-field addition to the SM that, in the first approximation, does not need to be embedded with care into an additional UV completion but instead necessitates only of some basic assumptions on the scalar potential. On the other hand, LQs bring their own bag of complications to the fixed-point analysis, which we will discuss in detail in the following sections.  First and foremost, they are not expected to be flavor-universal and there is in principle no knowledge of the alignment between the LQ and SM Yukawa matrices in flavor space. In practice, one can pick a particular basis for the fixed-point analysis and make sure that the basis remains well-defined along the entire RG flow. We will choose to work in the quark mass basis, in which the SM Yukawa matrices remain diagonal. As a direct consequence, we will treat the Cabibbo-Kobayashi-Maskawa (CKM) matrix parameters as renormalizable dimensionless couplings of the Lagrangian and add them to the fixed-point analysis as was done, e.g., in Ref.\cite{Alkofer:2020vtb}.  
 
 We will show that, if we require consistency with the neutral-current, $b\to s$ flavor anomalies, the information derived from the trans-Planckian fixed-point analysis leads to very specific predictions for the mass of the LQ, which should lie approximately in the $5-10\tev$ range. This places it outside of the foreseeable reach of the LHC, but well within the early reach of a 100\tev\ hadron machine according to the most conservative estimates. 
On the other hand, as the strength required for a NP contribution to the $b\to c$ anomalies is larger than in the neutral current case, 
our analysis confirms the well known fact that the charged-current anomalies point to
new states just above the current sensitivity of the LHC.
In this sense, one does not draw new indications from the UV completion, but we find it interesting \textit{per se} that this kind of solutions 
can be made consistent with a theoretical embedding so far unexplored in this context.  

The structure of the paper is the following. In \refsec{sec:asgrav} we give a brief overview of asymptotic safety 
in quantum gravity and we recall the mathematical setup used for the UV fixed-point analysis. In \refsec{sec:bsflav} we introduce the scalar LQ 
explanation for the $b\to s$ flavor anomalies. Subsections are dedicated to the full fixed-point analysis, a description of its solutions, and 
a summary of its low-scale predictions. In \refsec{sec:bcflav} we provide the full fixed-point analysis, description of solutions, and 
summary of low-scale predictions for the scalar LQ involved in the $b\to c$ case. 
We summarize our findings in \refsec{sec:summary}. Some technical details of the LQ models and of the RG flow analyses are given in Appendices~\ref{app:s3app} and \ref{app:s1app}.

\section{Asymptotic safety from quantum gravity}\label{sec:asgrav}

In the presence of non-negligible gravitational interactions, a regime expected to be entered while approaching the Planck scale, the RG flow of matter couplings is modified\cite{Robinson:2005fj,Pietrykowski:2006xy,Toms:2007sk,Tang:2008ah,Toms:2008dq,Rodigast:2009zj,Zanusso:2009bs,Daum:2009dn,Daum:2010bc,Folkerts:2011jz,Oda:2015sma,Eichhorn:2016esv,Christiansen:2017gtg,Hamada:2017rvn,Christiansen:2017cxa,Eichhorn:2017eht,Harst:2011zx,Eichhorn:2017lry}. For generic gauge ($g$) and Yukawa ($y$) couplings, such gravity-corrected beta functions are schematically given by
\bea
\beta_g&=&\beta_g^{\textrm{SM+NP}}-g\,f_g,\label{eq:asgrav}\\
\beta_y&=&\beta_y^{\textrm{SM+NP}}-y\,f_y,\label{eq:asgravfy}
\eea
where $\beta_x\equiv dx/d\log Q$, and the first term on the right hand side in both equations denotes standard contributions to the RG running 
from the SM and NP. The effect of gravitational interactions, captured by the parameters $f_g$ and $f_y$, is universal in the sense that gravity distinguishes between different types of matter interactions (gauge, Yukawa, scalar quartic, etc.), while, being blind to the internal symmetries, does not explicitly depend on the corresponding couplings. 

In the presence of asymptotically safe quantum gravity, the gravity-induced contributions $f_g$ and $f_y$ are determined by both the gravitational dynamics and the matter content of the coupled theory, whether this is the SM or its NP extension\cite{Zanusso:2009bs,Dona:2013qba,Eichhorn:2016esv,Eichhorn:2017eht,Christiansen:2017cxa}. While the SM and NP contributions are well defined in a given NP framework, the theoretical status of the parameters $f_g$ and $f_y$ is far from being definitively settled.

It was shown in Ref.\cite{Folkerts:2011jz} that the leading quantum gravity contribution to the 
gauge coupling beta functions cannot 
be negative, irrespective of a chosen RG scheme. In particular, one can prove that  
$f_g=0$ as long as the RG scheme preserves a certain symmetry of the classical gauge-gravity Lagrangian, whereas 
strictly positive values are obtained as soon as one breaks this symmetry, irrespective of any other technical choices.
Note, from \refeq{eq:asgrav}, that a strictly positive $f_g$ is required to enforce asymptotic freedom in the gauge sector, 
and in this sense one is inclined to choose an RG scheme in which the leading non-universal coefficient is non-zero to be 
consistent with the low-energy phenomenology. Conversely, higher-order calculations would be required to determine the  
fate of theories with $f_g=0$.
The existence of a non-trivial combined fixed point in a coupled system of gravity and matter has been confirmed in Ref.\cite{Christiansen:2017cxa}. It was shown there that in such a setup gravity is asymptotically safe, while the gauge sector asymptotically free (see also Ref.\cite{Litim:2011cp} for an early review). 

The theoretical status of the leading-order gravity correction $f_y$ is less clear. Only a set of simplified models has been analyzed in the literature in this context\cite{Rodigast:2009zj,Zanusso:2009bs,Oda:2015sma,Eichhorn:2016esv}, 
but no general results and definite conclusions regarding the sign of $f_y$ are available. Note, however, 
that quantum gravity is indispensable to generate a weakly coupled Yukawa fixed point\cite{Bond:2016dvk,Bond:2018oco}.

Additional unknowns adhere to the issue of how the gravity sector can carry the addition of matter, with large uncertainties that can stem from various sources. The first one is the choice of truncation of the theory space. In Einstein-Hilbert truncation only two operators in the scale-dependent effective action are retained, leading to the gravitational dynamics being governed exclusively by the Newton and cosmological constants\cite{Reuter:1996cp}. Inclusion of higher order interactions enriches the theory by additional free parameters\cite{Lauscher:2002sq,Codello:2007bd,Benedetti:2009rx,Falls:2017lst,Falls:2018ylp}. Secondly, within a chosen truncation, the derivation of gravity contributions to the matter beta functions is cutoff-scheme dependent\cite{Reuter:2001ag,Narain:2009qa} and various results can differ by up to 50-60\%\cite{Dona:2013qba}.

For all these reasons, we will follow the approach of Refs.\cite{Eichhorn:2017ylw,Eichhorn:2018whv,Reichert:2019car,Eichhorn:2020kca} and treat $f_g$ and $f_y$ as free parameters whose specific values will define a particular set of boundary conditions for the SM and  NP couplings  at the Planck scale, obtained by following the RG flow of the coupling system from the UV fixed point. On the other hand, the requirement of matching the SM parameters onto their experimentally measured values imposes strong limitations on the allowed magnitude of $f_g$ and $f_y$, where the former is usually determined by the low-energy hypercharge coupling.

A fixed point of the system of Eqs.~(\ref{eq:asgrav})-(\ref{eq:asgravfy}) is given by any set $\{g^\ast,y^\ast\}$, generically indicated with an asterisk, such that $\beta_g(g^\ast,y^\ast)=\beta_y(g^\ast,y^\ast)=0$.
In order to determine the structure of the fixed point one needs to analyze the RG flow in its vicinity. 
The standard method is to linearize the RG equation system of the couplings, $\{\alpha_i\}\equiv\{g,y\}$, around the fixed point. One derives the stability matrix, $M$,
\be\label{stab}
M_{ij}=\partial\beta_i/\partial\alpha_j|_{\{\alpha^{\ast}_i\}}\,,
\ee
whose eigenvalues $\theta_i$, called critical (or scaling) exponents, characterize the power-law evolution of the couplings in the vicinity of $\{\alpha^{\ast}_i\}$. 

If a critical exponent is negative the corresponding eigendirection is UV attractive and dubbed as {\it relevant}. All the RG trajectories along this direction will asymptotically reach the fixed point. A deviation of a relevant coupling from the fixed point introduces a free parameter in the theory and this freedom can be used to fine tune the coupling at some high scale to match an eventual measurement in the infrared~(IR). 
Conversely, if an eigenvalue of $M$ is positive, the corresponding eigendirection is UV repulsive and commonly dubbed as {\it irrelevant}. 
In this case there exist only one trajectory the coupling's flow can follow in its run to the IR, thus providing potentially a clear prediction for its value at the low, experimentally interesting scale. All the trajectories which emanate from the UV stable fixed points correspond to theories that remain finite at high energies. Finally, $\theta_i=0$ introduces a \textit{marginal} eigendirection. The RG flow along this direction is logarithmically slow and one needs to go beyond the linear approximation to decide whether a fixed point is attractive or repulsive.

\section{Flavor anomalies in $\boldsymbol{b\to s}$ transitions}\label{sec:bsflav}

We consider in this section the anomalies recorded in the last several years  
at LHCb\cite{Aaij:2014ora,Aaij:2014pli,Aaij:2015oid,Aaij:2015esa,Aaij:2016flj,Aaij:2017vbb,Aaij:2019wad,Aaij:2020nrf}, Belle\cite{Wehle:2016yoi,Abdesselam:2019wac}, CMS\cite{Sirunyan:2017dhj}, and ATLAS\cite{Aaboud:2018krd}, 
involving substantial deviations from the SM in the measured values 
of the lepton-flavor violating ratios and angular distributions of the decays $B\to K^{(\ast)}\mu\mu$.
Numerous global fits\cite{Altmannshofer:2014rta,Altmannshofer:2017fio,Capdevila:2017bsm,Altmannshofer:2017yso,DAmico:2017mtc,Ciuchini:2017mik,Alok:2017sui,Hurth:2014vma,Hurth:2016fbr,Chobanova:2017ghn,Hurth:2017hxg,Arbey:2018ics,Alguero:2019ptt,Alok:2019ufo,Ciuchini:2019usw,Datta:2019zca,Aebischer:2019mlg,Kowalska:2019ley,Arbey:2019duh,Bhattacharya:2019dot} 
have pointed to the likely emergence of NP in the effective operators 
\bea
\mathcal{O}^{\mu\,(\prime)}_9&=&\frac{\alpha_{\textrm{em}}}{4\pi}\left(\bar{s}\gamma_{\rho}P_{L(R)} b\right)\left(\bar{\mu}\gamma^{\rho}\mu\right)\,,\nonumber\\
\mathcal{O}^{\mu\,(\prime)}_{10}&=&\frac{\alpha_{\textrm{em}}}{4\pi}\left(\bar{s}\gamma_{\rho}P_{L(R)} b\right)(\bar{\mu}\gamma^{\rho}\gamma^5\mu)\,,
\eea
the statistical significance of which exceeds, according to some analyses, 
the $5\sigma$ level.

Among the NP scenarios well suited to induce the required deviation in the Wilson coefficient $C_9^{\mu},
$\footnote{A negative deviation from the SM expectation of $C_9^{\mu}$ 
alone is sufficient to explain the data, but other operators can be nonzero too, cf.~the global fits cited above.} 
LQs are particularly appealing for their simplicity. For example, 
the anomalies could be explained by the tree-level exchange of a single component of the 
scalar LQ $S_3$\cite{Hiller:2016kry,Dorsner:2016wpm,Dorsner:2017ufx,Crivellin:2017zlb,Hiller:2017bzc,Buttazzo:2017ixm,Hiller:2018wbv,Becirevic:2018afm,Angelescu:2018tyl}, which is a triplet of SU(2)$_L$. However, 
as is generally the case when matching renormalizable NP models to the flavor 
constraints expressed in terms of operators of dimension higher than~4, even a relatively precise measurement of the corresponding 
Wilson coefficient does not suffice to pinpoint the scale and interaction strength of the LQ independently, 
as the constraints are expressed in terms of a ratio coupling/mass. Thus, in order to make specific predictions, one has to introduce some assumptions on the nature of the UV completion. 
We show here that asymptotically safe gravity provides a predictive high-energy framework for the $S_3$ LQ in the context of the $b\to s$ anomalies.

We remind the reader that a solution to the flavor anomalies implies (we constrain ourselves for simplicity to
the single dimension $C_9^{\mu}=-C_{10}^{\mu}$)
\be
C_9^{\mu}=-C_{10}^{\mu}\in \left( -0.7,-0.3\right)\label{eq:bsconst}
\ee 
at the $2\sigma$ level\cite{Kowalska:2019ley}. 
The corresponding operator can be generated by the LQ $S_3$, whose SU(3)$_c\times$SU(2)$_L\times$U(1)$_Y$ quantum numbers are $(\mathbf{\bar{3}},\mathbf{3},1/3)$. We can write down the interaction Lagrangian in terms of left-chiral two-component spinors in the mass basis (we use the Weyl notation throughout this work),
\begin{multline}
\mathcal{L}\supset \hat{Y}^L_{ij}\left(-\phi_{1/3}d_{L,i}\nu_{L,j}-\sqrt{2}\phi_{4/3}d_{L,i}e_{L,j}\right)\\
+\widetilde{Y}^L_{ij}\left(\sqrt{2}\phi_{-2/3}u_{L,i}\nu_{L,j}-\phi_{1/3}u_{L,i}e_{L,j}\right)+\textrm{H.c.}\,,
\end{multline}
where a sum over repeated SM generation indices is intended, numbers in subscripts indicate the scalar fields' electric charge, 
and the up- and down-type couplings are 
related to each other 
by the CKM matrix $V$ as
$\widetilde{Y}_{ij}=V^{\ast}_{ik}\hat{Y}_{kj}$\,. 
Note that we do not address in this work the physics of neutrinos, which we assume form a separate system not affecting the fixed-point analysis of states that are heavier by several orders of magnitude. We thus constrain ourselves to a SM-like framework, 
in which the charged lepton generations do not mix with one another. 
Additional details on the $S_3$ Lagrangian and the complete list of adopted assumptions are given in Appendix~\ref{app:s3app}.

By matching to the $\mathcal{O}_9^{\mu}$ operator via the $t$-channel exchange of $\phi_{4/3}$ one gets
\be\label{eq:c910}
C_9^{\mu}=-C_{10}^{\mu}=\frac{\pi v_h^2}{V_{33}V^{\ast}_{32}\alpha_{\textrm{em}}}\frac{\hat{Y}^L_{32}\hat{Y}^{L\ast}_{22}}{m_{S_3}^2}\,,
\ee
where $v_h$ is the Higgs vev, $\alpha_{\textrm{em}}$ is the fine structure constant, and 
we have attributed a common mass $m_{S_3}$ to the triplet's states.  

Equation~(\ref{eq:bsconst}) leads to the $2\sigma$ bound
\be\label{eq:bsbound}
0.4\times 10^{-3} \left(\frac{m_{S_3}}{\tev}\right)^2 \leq\hat{Y}^L_{32}\hat{Y}_{22}^{L\ast}\leq 1.1\times 10^{-3} \left(\frac{m_{S_3}}{\tev}\right)^2\,,
\ee
which, as was explained above, does not provide an independent determination of the LQ mass and coupling. 
On the other hand, if a specific value of the Yukawa couplings were to emerge from the UV completion, 
\refeq{eq:bsbound} would provide a clear indication of the LQ mass scale, subject only to the experimental
precision of the flavor measurements.
In the next subsection we derive the value 
of the LQ Yukawa coupling from the fixed-point analysis of the system in minimally coupled quantum gravity 
in the trans-Planckian regime. 

\subsection{Fixed-point analysis}

Since we seek to connect the high-scale boundary conditions with observable quantities at the low scale, we derive the renormalization group equations (RGEs) in the quark mass basis. Besides, since we use specifically the measurements of $b\to s$ anomalies to constrain the NP 
system at the low scale, we choose to work in a down-origin basis for the $S_3$ Yukawa matrices. A more detailed discussion of our choice of basis can be found in Appendix~\ref{app:s3app}.  

As the Yukawa matrices of the SM and those of the LQ system do not necessarily commute, we are left with diagonal Yukawa entries for the SM
and arbitrary textures for the down-type $\hat{Y}^L_{ij}$ matrices. However, since we neglect the neutrino mass, 
we can always choose a charged-lepton Yukawa matrix diagonal in flavor space and thus we do not generate inter-column, charged lepton-flavor violating elements via RG flow. The CKM matrix elements are subject to RG running like the SM and NP Yukawa couplings. Above the Planck scale, the system is coupled to the quantum fluctuations of the graviton, which introduce the $f_g$ and $f_y$ terms in the RGEs and give rise to the possible emergence of UV (Gaussian and interactive)
fixed points. 

In the present analysis we focus only on the parameters affecting the phenomenological $b\to s$ constraints. This means that we work effectively in a 2-family (second + third) approximation in which the CKM matrix is parametrized by one single rotation angle. 
Additionally, we do not include in the analysis the Yukawa couplings of the quarks of the first two generations, as their impact on the running of other parameters is negligible. 
One should expect to be able to set these negligible parameters along a relevant direction of a Gaussian fixed point in the trans-Planckian UV\cite{Alkofer:2020vtb}.

The minimal system of couplings consists of 8 independent parameters,
\be\label{S3_par}
g_Y,\,g_2,\,g_3,\,y_t,\,y_b,\,\hat{Y}^L_{22},\,\hat{Y}^L_{32},\,V_{33},
\ee
where, respectively, $g_Y$, $g_2$ and $g_3$ are the gauge couplings of U(1)$_Y$, SU(2)$_L$ and SU(3)$_c$, and $y_t$ and $y_b$ denote the Yukawa couplings of top and bottom quarks. Note that the Yukawa coupling $\hat{Y}^L_{12}$ does not enter the fixed-point analysis
in our approximation. We do make sure, however, that if it is assumed to be zero at the Planck scale, it does not get renormalized at the low scale into values in tension with the experimental bounds.
Finally, we point out that we limit our analysis to real couplings only. The relevant RGEs for the $S_3$ plus SM system coupled to gravity are given in Appendix~\ref{app:s3app}.\footnote{It is 
worth pointing out 
that we do not incorporate the parameters of the scalar potential in the fixed-point analysis. They do not enter at one loop in the RGEs of the gauge-Yukawa system, 
for which we derive the phenomenological predictions related to the flavor anomalies, see Appendix~\ref{app:s3app}.
Moreover, we have checked numerically that the predictions for the NP Yukawa couplings change minimally under the addition of perturbative 2-loop contributions to the beta functions. This variation is 
negligible with respect to the experimental uncertainty on the Wilson coefficients and does 
not affect the leptoquark mass determination. }

Let us explore the structure of the fixed points for the system given by \refeq{S3_par}. The non-abelian gauge couplings develop non-interactive UV fixed points, indicated henceforth with an asterisk, 
\be
g_3^\ast=0,\qquad g_2^\ast=0,
\ee
therefore they will correspond to the relevant directions in the coupling space. The trans-Planckian running of $g_Y$, on the other hand, is tamed by the graviton fluctuations, which lead to the generation of an interactive fixed point. As discussed in \refsec{sec:asgrav}, matching the irrelevant $g_Y$ onto its phenomenological value in the IR 
allows one to unambiguously fix the parameter $f_g$,
\be
g_Y^\ast=4\pi\sqrt{\frac{6\,f_g}{43}}\,.
\ee
One obtains $f_g=0.01$ and $g_Y^\ast=0.48$.
 
The second quantum gravity parameter, $f_y$, can be fixed if one of the SM Yukawa couplings presents a UV interactive fixed point\cite{Eichhorn:2018whv}, as in that case it is used to match the flow to the IR along the irrelevant direction onto the value of the corresponding quarks mass. At the same time, the CKM matrix element $V_{33}$ must be set to zero at the fixed point to span a relevant direction\cite{Alkofer:2020vtb}. The following combinations of the fixed-point values in the SM Yukawa sector are then possible:
\bea
&\textrm{FP}_1:& y_t^\ast\neq 0, \quad y_b^\ast= 0, \quad  V_{33}^\ast= 0,\nonumber \\
&\textrm{FP}_2:& y_t^\ast= 0, \quad y_b^\ast\neq 0, \quad V_{33}^\ast= 0,\nonumber\\
&\textrm{FP}_3:& y_t^\ast\neq 0, \quad y_b^\ast\neq 0, \quad V_{33}^\ast= 0.
\eea

Finally, three different sets of fixed-point values of the LQ Yukawa matrix entries can be obtained, 
allowing one of the two elements $ \hat{Y}^L_{22}$, $\hat{Y}^L_{32}$ to be interactive, or none:
\bea\label{mod:bsm}
&\textrm{FP}_a:& \hat{Y}^{L\ast}_{22}\neq 0,\;\hat{Y}^{L\ast}_{32}= 0,\nonumber \\
&\textrm{FP}_b:& \hat{Y}^{L\ast}_{22}= 0,\;\hat{Y}^{L\ast}_{32}\neq 0,\nonumber \\
&\textrm{FP}_c:& \hat{Y}^{L\ast}_{22}= 0,\;\hat{Y}^{L\ast}_{32}= 0.
\eea

Note that a solution with both non-zero elements of the matrix $\hat{Y}_{ij}^L$ is not compatible with a relevant fixed point for $V_{33}$.
The list of fixed points of phenomenological interest is summarized in \reftable{tab:S3FP}. 
The values assumed by $f_y$ at the various fixed points are also presented in \reftable{tab:S3FP}.

\begin{table*}[t]
\footnotesize
\begin{center}
\begin{tabular}{|c|c|cccc|c|}
\hline
 &$f_y$ &$y_t^\ast$ & $y_b^\ast$ & $\hat{Y}_{22}^{L\ast}$ & $\hat{Y}_{32}^{L\ast}$ & Prediction \\
\hline
$\textrm{FP}_{1a}$ & 0.0014 & $4\pi \sqrt{\frac{208 f_g + 946 f_y}{ 5289}}$& 0 & $4\pi \sqrt{\frac{146 f_g + 1376 f_y}{5289}}$ & 0 & $\hat{Y}_{22}^L\hat{Y}_{32}^L(m_{S_3})<0$ \\
$\textrm{FP}_{1b}$ & 0.0088 & $4\pi \sqrt{\frac{17 f_g + 86 f_y}{387}}$& 0 & 0 & $4\pi \sqrt{\frac{10 f_g + 86 f_y}{301}}$ & $ m_{S_3}\in [7, 11]\tev\,^{\dagger}$ \\
$\textrm{FP}_{1c}$ & 0.0024 & $4\pi \sqrt{\frac{17 f_g + 86 f_y}{387}}$ & 0 & 0 & 0 & $ m_{S_3}\in [7, 11]\tev\,^{\dagger}$\\
\hline
$\textrm{FP}_{2a}$ & -0.0006 & 0& $4\pi \sqrt{\frac{5 f_g + 86 f_y}{387}}$& $4\pi \sqrt{\frac{10 f_g +86f_y}{301}}$ & 0 & $\hat{Y}_{22}^L\hat{Y}_{32}^L(m_{S_3})<0$\\
$\textrm{FP}_{2b}$ & -0.0004 & 0& $4\pi\sqrt{\frac{40f_g + 946 f_y}{5289}}$& 0 & $4\pi \sqrt{\frac{170 f_g + 1376 f_y}{5289}}$ &  $ m_{S_3}\in [4, 7]\tev$ \\
$\textrm{FP}_{2c}$ &  -0.0001  & 0& $4\pi\sqrt{\frac{5f_g + 86 f_y}{387}}$& 0 & 0 & $ m_{S_3}\in [4, 7]\tev$  \\
\hline
$\textrm{FP}_{3a}$ & 0.0014 & $4\pi \sqrt{\frac{484 f_g + 430 f_y}{8643}}$& $4\pi \sqrt{\frac{-211 f_g + 1634 f_y}{8643}}$& $4\pi \sqrt{\frac{218 f_g +2408f_y}{8643}}$ & 0 & $\hat{Y}_{22}^L\hat{Y}_{32}^L(m_{S_3})<0$\\
$\textrm{FP}_{3b}$ &0.0087 & $4\pi \sqrt{\frac{617 f_g + 1634 f_y}{8643}}$ & $4\pi \sqrt{\frac{-356f_g + 430 f_y}{8643}}$& 0 & $4\pi \sqrt{\frac{338 f_g + 2408 f_y}{8643}}$ &  $ m_{S_3}\in [7, 11]\tev\,^{\dagger}$\\
$\textrm{FP}_{3c}$ & 0.0023 & $4\pi \sqrt{\frac{41 f_g + 86 f_y}{645}}$ & $4\pi \sqrt{\frac{-19 f_g + 86 f_y}{645}}$& 0 & 0 &  $ m_{S_3}\in [7, 11]\tev\,^{\dagger}$\\
\hline
\end{tabular}
\caption{Fixed-point values and the corresponding $f_y$ of the SM+$S_3$ Yukawa coupling system invoked 
for an explanation of the $b\to s$ anomalies. The low-energy prediction for the NP sector is shown in the last column on the right. 
The symbol $\dag$ indicates those scenarios in which the predicted top mass exceeds the experimental value.}
\label{tab:S3FP}
\end{center}
\end{table*}

The SM couplings correspond directly to eigendirections of the stability matrix. 
The relevant couplings are $g_3$, $g_2$, and the one among the top and bottom Yukawa couplings whose fixed-point value is zero, 
denoted with $y_0$ in what follows. 
The deviation of the relevant couplings from their UV fixed point at some high scale $\Lambda$ introduces several free parameters characterizing UV-safe trajectories running out of the fixed point:
\be
\delta g_{2,3}(\Lambda)=g_{2,3}^\ast-g_{2,3}(\Lambda),\qquad \delta y_0=y_0^\ast-y_0(\Lambda).
\ee
Conversely, the irrelevant couplings $g_Y$ and $y_{\slashed{0}}$ (the SM Yukawa coupling(s) whose fixed-point value is nonzero) are expected to be entirely determined by their fixed-point value and thus constitute predictions of the theory. 
The behavior of the NP Yukawa couplings in the vicinity of the fixed point depends on which of the scenarios introduced in \refeq{mod:bsm} is considered and will be discussed individually in the following paragraphs.

A word of caution is in order, though. The character of a coupling as a relevant parameter does not need 
to persist along its entire RG flow from the UV to the IR, as it may be affected by the existence of other fixed points in the system.
In this regard, we anticipate here that all of the considered solutions found for the SM+$S_3$ leptoquark system eventually cross over in their trans-Planckian flow to the basin of attraction of a fixed point characterized by IR-attractive $V_{33}^\ast=1$, $y_t^\ast\neq 0$, $y_b^\ast = 0$, $\hat{Y}^{L\ast}_{22}\neq 0$ and $\hat{Y}^{L\ast}_{32}= 0$. Explicitly,
\be\label{eq:f1til}
y_t^\ast=\frac{4\pi}{3} \sqrt{\frac{17 f_g + 86f_y}{ 43}},\qquad \hat{Y}^{L\ast}_{22}=4\pi \sqrt{\frac{10 f_g + 86f_y}{301}}\,.
\ee

We dub this trans-Planckian IR fixed point as FP$_{\textrm{IR}}$. Its origin can be qualitatively understood by inspection of the RGEs
presented at the end of Appendix~\ref{app:s3app}. Equations~(\ref{eq:yt_S3}) and (\ref{eq:L22_S3}) imply 
that, because of the negative contributions induced by the hypercharge coupling, both $y_t$ and $\hat{Y}^{L}_{22}$ can eventually become 
substantial in their flow to the IR, independently of their starting point in the UV. The consequence of this growth is that
the contribution proportional to $g_Y^2$ will be counterbalanced by $y_t^2$ in \refeq{eq:yt_S3} and by $(\hat{Y}^{L}_{22})^2$ in
\refeq{eq:L22_S3} (with some correction proportional to $y_t\,\hat{Y}^{L}_{32}$ that kicks in when $V_{33}^2\approx V_{32}^2$). 
This effect is dominated by the
size of $g_Y^{\ast}$ and hence by the gravitational coupling $f_g$, and it is much less sensitive to the parameter $f_y$.
As we shall see in the next paragraphs, in most of the cases of phenomenological interest the presence of FP$_{\textrm{IR}}$ along the RGE flow effectively washes out much of the freedom associated with relevant directions in the Yukawa-coupling theory space.

 \begin{figure*}[t]
	\centering
	\subfloat[FP$_{1a}$]{%
		\includegraphics[width=0.4\textwidth]{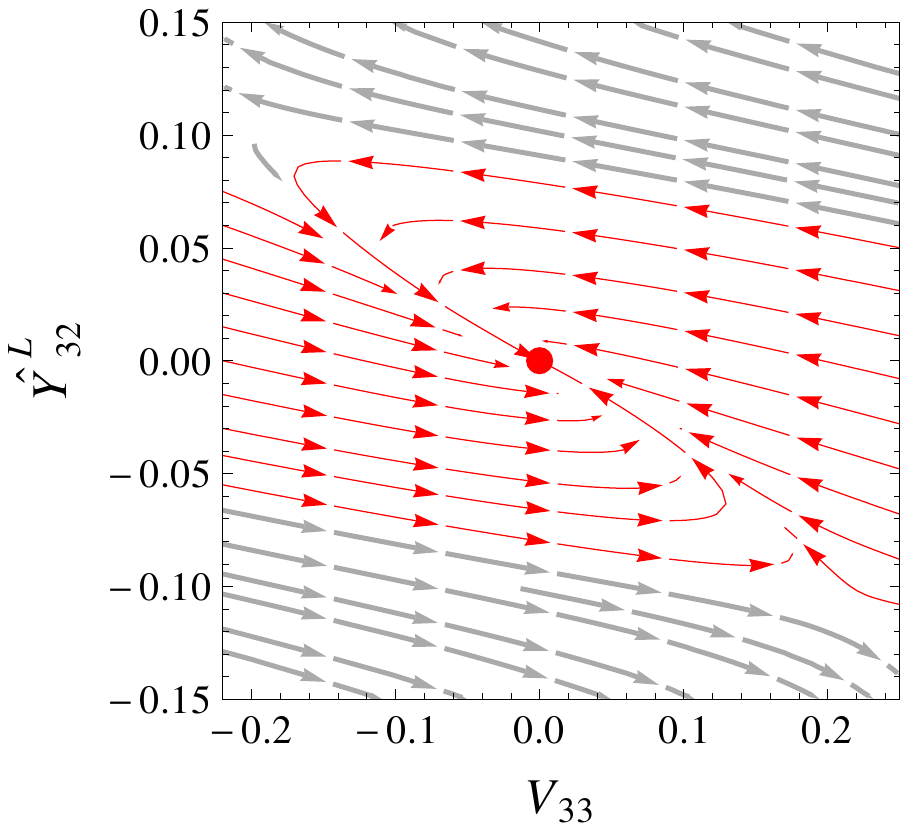}
	}%
	\hspace{0.8cm}
		\subfloat[FP$_{1b}$]{%
		\includegraphics[width=0.4\textwidth]{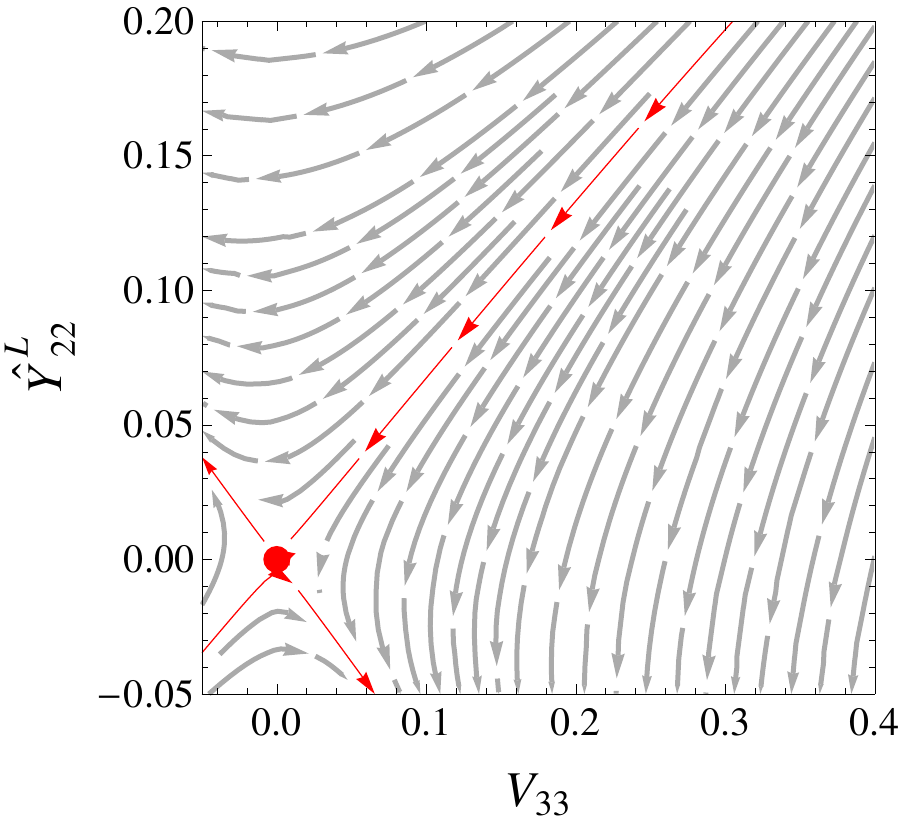}
	}%
	\caption{(a) Phase diagram in the plane ($V_{33}$, $\hat{Y}^L_{32}$) in the vicinity of the UV fixed point $\textrm{FP}_{1a}$, indicated here as a red dot. All the remaining couplings are set to their fixed-point values. The RG flow directions point towards the UV. Only the RG trajectories marked in red allow one to reach the UV fixed point. (b) Phase diagram in the plane ($V_{33}$, $\hat{Y}^L_{22}$) in the vicinity of the UV fixed point $\textrm{FP}_{1b}$.}
			\label{fig:stream_S3a}
\end{figure*}

\paragraph{Scenarios of type FP$_a$} For a fixed point of the type FP$_{1a}$ or FP$_{3a}$, the coupling $\hat{Y}^L_{22}$ is irrelevant, 
while the system ($V_{33}$, $\hat{Y}^L_{32}$) spans a 2-dimensional submanifold in the coupling space, 
on which both couplings are relevant but do 
not correspond to eigendirections of the stability matrix. 

The phase diagram of $\hat{Y}^L_{32}$ vs.~$V_{33}$ in the vicinity of the fixed point is shown in \reffig{fig:stream_S3a}(a), with all the
remaining couplings fixed at their fixed-point value. 
The fixed point is marked as a red dot, and the arrows indicate the RGE flow of the system towards the UV. 
For the trajectories shown in red, the fixed point can be reached from any direction, confirming that both couplings are indeed 
relevant.\footnote{Note in \reffig{fig:stream_S3a}(a), that the streamlines flow with different speed along the $V_{33}$ and $\hat{Y}^L_{32}$ 
directions, giving the impression of entering the fixed point in the UV along one and the same line.} However, for any fixed deviation $\delta V_{33}$, there exists an upper bound on the allowed size of the corresponding $\delta \hat{Y}^L_{32} $, and \textit{viceversa}, which is due 
to the nontrivial beta function of the element $V_{33}$. 
This fact bears the important consequence that $\delta \hat{Y}^L_{32}$ 
never reaches a value large enough to guarantee that 
$\hat{Y}^L_{32}$ remain positive along the full length of its flow to the IR. More specifically, 
close to the relevant fixed point the running of $\hat{Y}^L_{32}$ is dominated by $y_t$, which introduces a positive contribution to the beta function (see \refeq{eq:L32_S3} in Appendix~\ref{app:s3app}), 
\be\label{eq:runY32}
\frac{d\hat{Y}_{32}^L}{dt}\simeq \frac{1}{16\pi^2}\hat{Y}_{22}^L\,y_t^2\,V_{33}\sqrt{1-V_{33}{}^2}\,.
\ee
Inevitably, in scenarios FP$_{1a}$, FP$_{3a}$ one obtains $\hat{Y}_{22}^L\hat{Y}_{32}^L<0$ at the low scale, in contradiction with the phenomenological requirement, cf.~Eqs.~(\ref{eq:bsconst}), (\ref{eq:c910}).

The situation is very similar for fixed point $\textrm{FP}_{2a}$, although in this case $\hat{Y}^L_{32}$ 
corresponds to an irrelevant direction. Once more, due to \refeq{eq:runY32}, the low-scale prediction is phenomenologically disfavored.

\paragraph{Scenarios of type FP$_b$} In this case both LQ Yukawa couplings become irrelevant directions. While $\hat{Y}_{32}^L$ corresponds to an eigenvector of the stability matrix, the flow of $\hat{Y}_{22}^L$ close to the fixed point is entirely dictated by the UV hypercritical surface relating it with the relevant CKM matrix element $V_{33}$: $\hat{Y}^L_{22}(t)\equiv \mathcal{F}(V_{33}(t))$.
The corresponding phase diagram is shown in \reffig{fig:stream_S3a}(b). 
Since in this case the main contribution to the $\hat{Y}^L_{22}$ beta function is negative (see \refeq{eq:L22_S3} in Appendix~\ref{app:s3app}),
\be\label{eq:runY22}
\frac{d\hat{Y}_{22}^L}{dt}\simeq -\frac{2}{16\pi^2}\hat{Y}_{32}^L\,y_t^2\,V_{33}\sqrt{1-V_{33}{}^2},
\ee
the product $\hat{Y}_{22}^L\hat{Y}_{32}^L$ is positive at the low scale, as required by global fits to the Wilson coefficients, 
$C_9^{\mu}=-C_{10}^{\mu}<0$.

The phenomenological predictions, obtained by following the RG flow of the examined coupling system from the 
UV fixed point towards the IR, differ somewhat for FP$_{1b}$, FP$_{2b}$, and FP$_{3b}$. 
In all three cases the requirement to fit the measured value of the hypercharge coupling, which we assume to be $g_Y(M_t)=0.3583$\cite{Buttazzo:2013uya}, fixes $f_g=0.01$.
Case~$\textrm{FP}_{1b}$ features irrelevant nonzero top Yukawa coupling at the  fixed point. But whether $y_b^{\ast}=0$ corresponds instead to a relevant or irrelevant direction -- a key question from the point of view of the phenomenological viability of this scenario -- hinges on the precise value of $f_y$. For small $f_y$, $y_b^{\ast}=0$ is irrelevant. Thus, 
we choose a minimal $f_y$ for which $y_b$ becomes relevant: $f_y=0.0088$. 
One then obtains the following low-energy predictions for the SM and LQ Yukawa couplings at $Q=m_{S_3}=9\tev$,
\bea\label{eq:FP1b}
 y_t(m_{S_3})=1.07, & &y_b(m_{S_3})=0.01,\nonumber\\
 \hat{Y}^L_{22}(m_{S_3})=1.04,& & \hat{Y}^L_{32}(m_{S_3})=0.05\,.
\eea
The top mass is by around 20\% too large with respect to the SM predictions -- a price to pay for fitting correctly the bottom mass in scenario~$\textrm{FP}_{1b}$. 

 \begin{figure*}[t]
	\centering
	\subfloat[FP$_{1b}$]{%
		\includegraphics[width=0.4\textwidth]{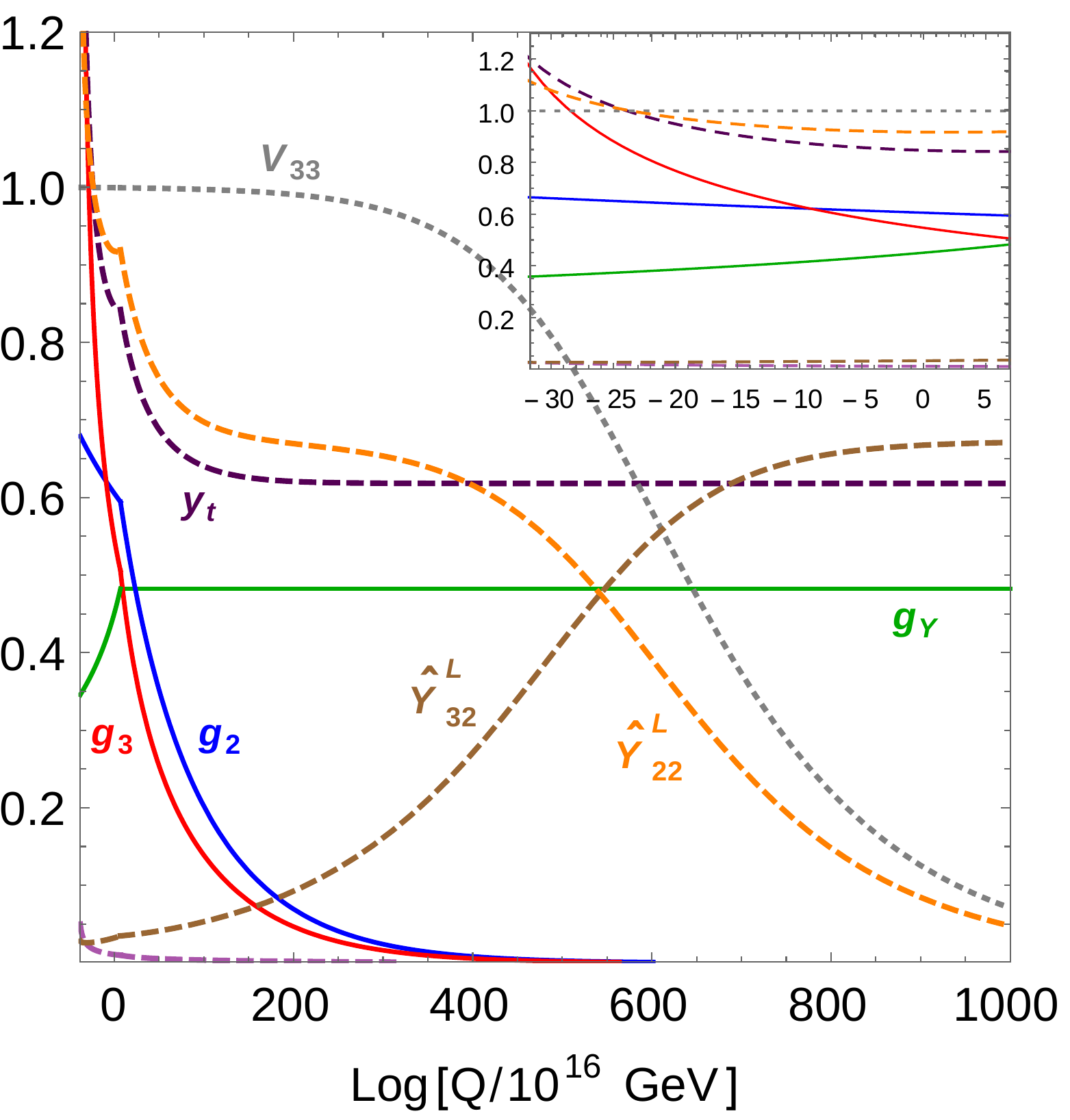}
	}%
	\hspace{0.8cm}
		\subfloat[FP$_{2b}$]{%
		\includegraphics[width=0.405\textwidth]{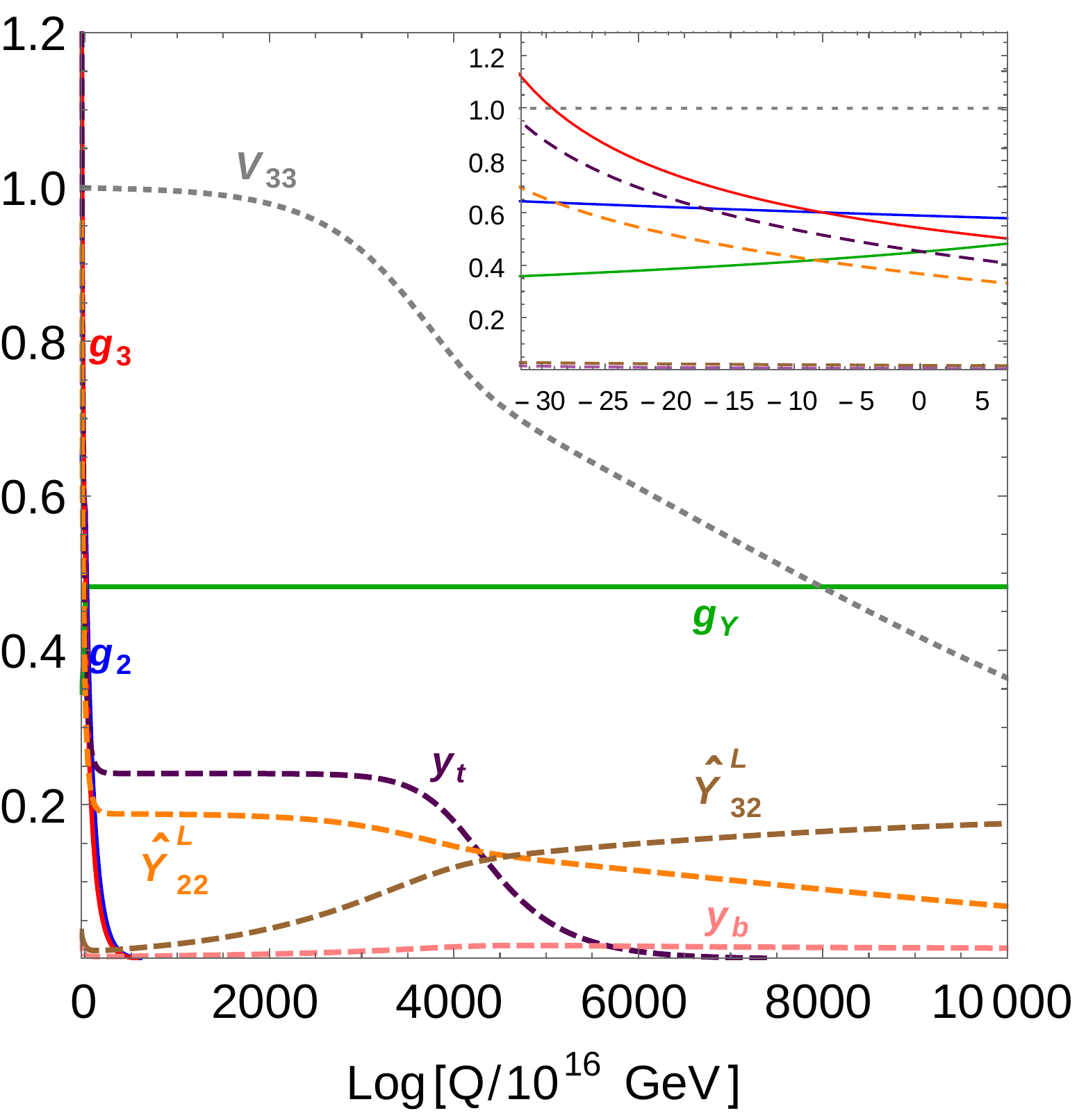}
	}%
	\caption{RG flow of the gauge and Yukawa couplings from the trans-Planckian energies down to the EWSB scale for scenarios featuring the fixed point (a) FP$_{1b}$, and (b) FP$_{2b}$. The sub-Planckian flow is depicted in the inset panels.}
			\label{fig:run1b}
\end{figure*}

The trans-Planckian flow of the parameters of the system is presented in \reffig{fig:run1b}(a). 
Their IR behavior is determined by the crossover towards the basin of attraction of fixed point FP$_{\textrm{IR}}$, 
which happens around $\log Q\approx 300$ in \reffig{fig:run1b}(a). FP$_{\textrm{IR}}$ 
is characterized by IR-attractive $V_{33}^\ast=1$, 
$y_t^\ast\approx 0.62$, $y_b^\ast = 0$, and $\hat{Y}^{L\ast}_{22}\approx 0.67$, cf.~\refeq{eq:f1til}. Since in this case  $\hat{Y}_{32}^L$
is large and positive in the deep UV, it remains positive in its flow towards the IR while asymptotically approaching zero from above.

Given the fixed values of $f_g$ and $f_y$, the EWSB values of $\hat{Y}^L_{22}$, and $\hat{Y}^L_{32}$ are unambiguously predicted by the RG flow towards the IR after gravity decouples (see inset panel in \reffig{fig:run1b}(a)). Moreover, since both LQ couplings run slowly over several orders of magnitudes in scale, one can precisely determine the range of LQ masses for which the analyzed scenario is consistent with the explanation of $b\to s$ anomalies. It corresponds to $m_{S_3}= 7-11\tev$.

Fixed point $\textrm{FP}_{2b}$ features a different behavior. The parameter $f_y$ is precisely determined by fitting $y_b$ to its IR value giving: $f_y=-0.0004$. At $Q=m_{S_3}=5\tev$ we obtain 
\bea\label{eq:FP2b}
 y_t(m_{S_3})=0.83,& & y_b(m_{S_3})=0.01,\nonumber\\
 \hat{Y}^L_{22}(m_{S_3})=0.63, & & \hat{Y}^L_{32}(m_{S_3})=0.03\,.
\eea
Note that the top mass can be matched onto the SM with a high degree of accuracy.
The corresponding range of LQ mass reads $m_{S_3}= 4-7\tev$. 
The trans-Planckian flow of the parameters of the system is presented in \reffig{fig:run1b}(b). At the energies 
$Q\approx \exp(4000)\approx 10^{1700}\gev$ 
one observes the crossover of the top Yukawa coupling from the basin of attraction of the UV fixed point FP$_{2b}$ to the basin of attraction of its IR fixed point FP$_{\textrm{IR}}$ with $y_t^\ast\approx 0.24$. 
It results in a characteristic ``plateau'' in the running of $y_t$ and allows for a good fit to the top 
mass once gravitational interactions decouple.

Finally, $\textrm{FP}_{3b}$ behaves quite similarly to $\textrm{FP}_{1b}$\,.
While $f_y$ is fixed by the bottom mass, the predicted mass of the top quark at low-energies is once more a little too large. 
At $Q=m_{S_3}=9\tev$ we obtain 
\bea\label{eq:FP3b}
 y_t(m_{S_3})=1.07,& & y_b(m_{S_3})=0.013,\nonumber\\
 \hat{Y}^L_{22}(m_{S_3})=1.04,& & \hat{Y}^L_{32}(m_{S_3})=0.05\,.
\eea

Note that there is a difference between cases FP$_{1b}$, FP$_{3b}$, where $y_t$ is irrelevant with $f_y=0.0088$,  and FP$_{2b}$, where it is relevant with $f_y=-0.00044$. The size of $f_y$ determines the size of the top Yukawa coupling at FP$_{\textrm{IR}}$, 
which reads 0.62 and 0.24, respectively. In the latter case, $|f_y| \ll f_g$, similarly to one of the SM cases presented in 
Ref.\cite{Alkofer:2020vtb}, and this leads to a good top mass determination.

 \begin{figure*}[t]
	\centering
	\subfloat[FP$_{1c}$]{%
		\includegraphics[width=0.4\textwidth]{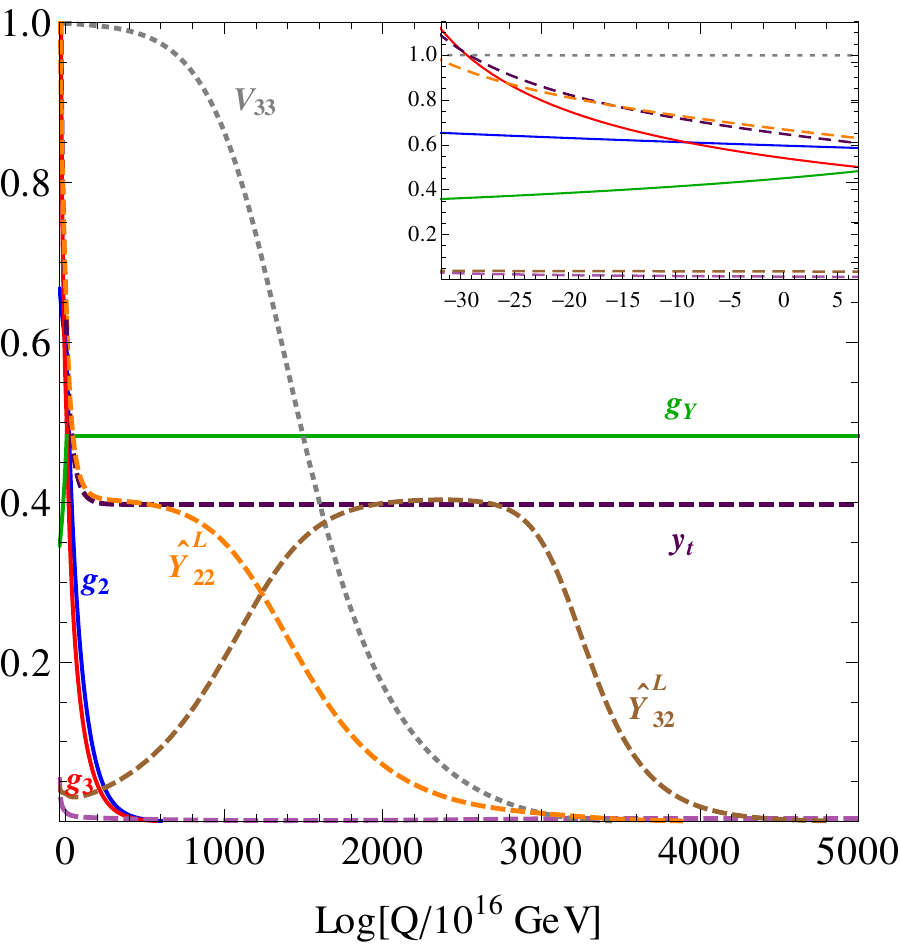}
	}%
	\hspace{0.8cm}
		\subfloat[FP$_{2c}$]{%
		\includegraphics[width=0.405\textwidth]{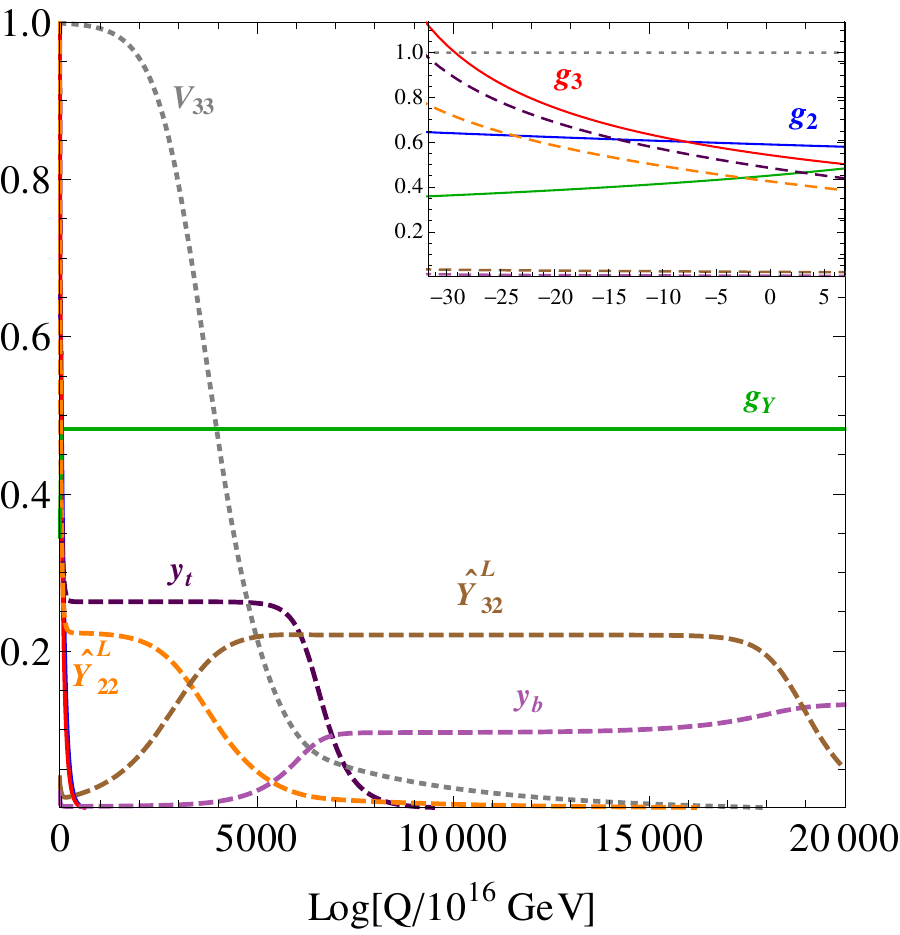}
	}%
	\caption{RG flow of the gauge and Yukawa couplings from the trans-Planckian energies down to the EWSB scale for scenarios featuring the fixed point (a) FP$_{1c}$, and (b) FP$_{2c}$. The sub-Planckian flow is depicted in the inset panels.}
			\label{fig:runc}
\end{figure*}

\paragraph{Scenarios of type FP$_c$} Fixed points of type FP$_c$ are characterized by two relevant directions for the Gaussian 
$\hat{Y}^{L\ast}_{22}=0$ and $\hat{Y}^{L\ast}_{32}=0$. However, predictivity is restored in these cases by virtue of the flow of the CKM matrix element $V_{33}$. Matching it to its low-energy experimental determination once again drives the system, first 
to the basin of attraction of fixed points reminiscent of type FP$_b$, and subsequently to the basin of attraction of FP$_{\textrm{IR}}$. The latter determines the values of  $\hat{Y}^{L}_{22}$ and 
$\hat{Y}^{L}_{32}$ at the Planck scale. The low-scale predictions for 
FP$_{1c}$, FP$_{2c}$, and  FP$_{3c}$ resemble closely the predictions for 
FP$_{1b}$, FP$_{2b}$, and  FP$_{3b}$, with the former and latter featuring a slightly too large top quark mass (in both cases $y_t$ emerges from an irrelevant UV fixed point), and the middle one instead providing a good fit to the top mass ($y_t^{\ast}=0$ is 
relevant in the UV) and NP predictions closely aligned with those of FP$_{2b}$. The trans-Planckian flow for case FP$_{1c}$ is depicted in \reffig{fig:runc}(a), whereas the one for FP$_{2c}$ is shown in \reffig{fig:runc}(b). Note the presence of two subsequent 
IR fixed points: at the center of the plots one can see the cross-over towards the FP$_b$-like fixed points, whereas 
close to the Planck scale towards FP$_{\textrm{IR}}$.

 \begin{figure*}[t]
	\centering
		\includegraphics[width=0.50\textwidth]{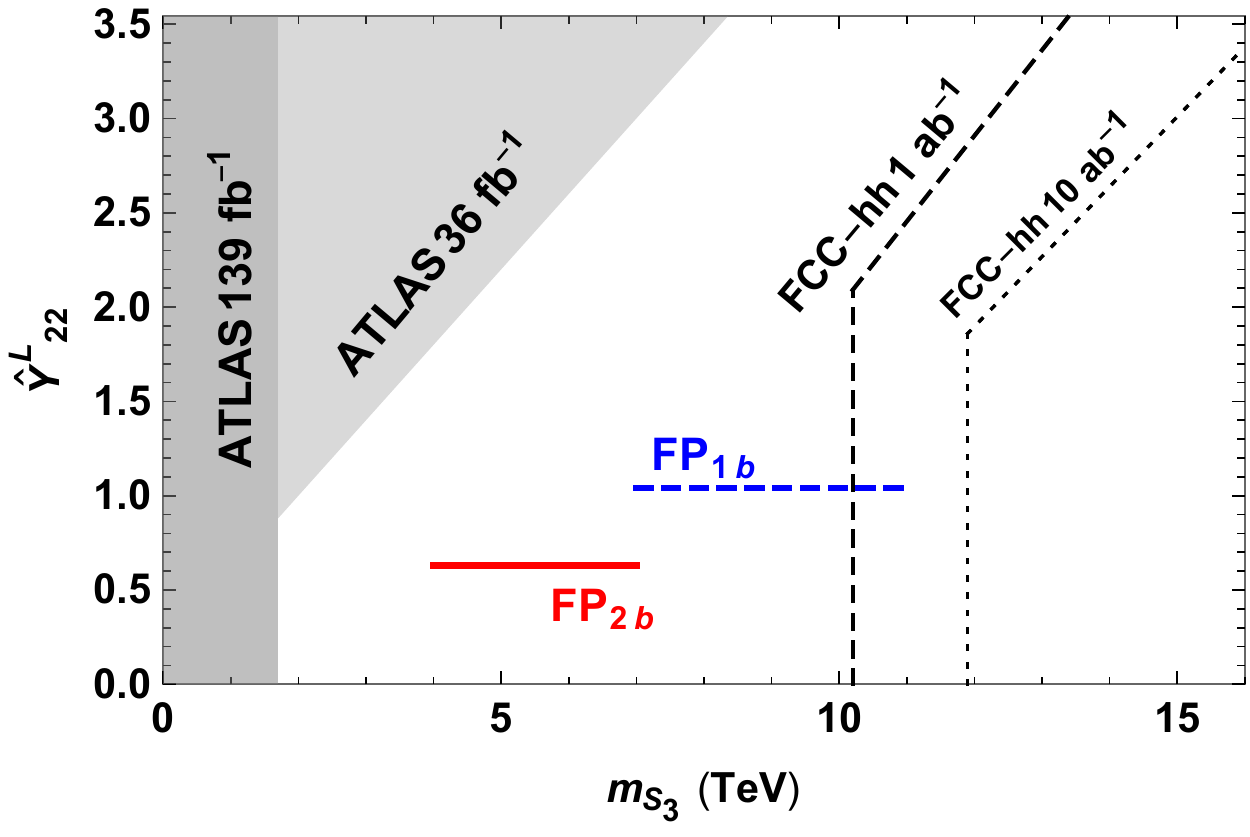}
	\caption{The low-energy Yukawa coupling and mass ranges predicted by the trans-Planckian fixed-point analysis 
	for the LQ $S_3$ to be in $2\sigma$ agreement with the $b\to s$ anomalies. 
	In red solid the range for FP$_{2b}$, in dashed blue the range for FP$_{1b}$. 
	Dark gray region shows the current lower bound from color LQ pair-production at ATLAS\cite{Aad:2020iuy}, light gray region gives the bound from single quark production as recast in Ref.\cite{Angelescu:2018tyl}. The dashed (dotted) black line marks the estimated reach of the hadron-hadron collider FCC-hh at $\sqrt{s}=100\tev$ and $1\,\textrm{ab}^{-1}$ ($10\,\textrm{ab}^{-1}$) integrated luminosity\cite{Abada:2019lih}.}
			\label{fig:FCC_S3}
\end{figure*}

\subsection{Low-scale predictions}

We show in \reffig{fig:FCC_S3} the LQ mass and coupling position in the plane $(m_{S_3},\hat{Y}^L_{22})$, predicted to be in $2\sigma$ agreement\cite{Kowalska:2019ley} with the $b\to s$ anomalies in FP$_{1b}$ (dashed blue) and FP$_{2b}$ (solid red). 
The case FP$_{3b}$ roughly overlaps with FP$_{1b}$ and is not shown in the plot. The same can be said for FP$_{1c}$, FP$_{3c}$, which roughly overlap with FP$_{1b}$, and for FP$_{2c}$, which overlaps with FP$_{2b}$.
In dark gray we show the current lower bound from LQ pair production at ATLAS\cite{Aad:2020iuy}, and in light gray the bound from single quark production as recast in Ref.\cite{Angelescu:2018tyl}. The dashed (dotted) line marks the estimated reach of the hadron-hadron collider FCC-hh at $\sqrt{s}=100\tev$ and $1\,\textrm{ab}^{-1}$ ($10\,\textrm{ab}^{-1}$) integrated luminosity\cite{Abada:2019lih}.
While the predicted LQ mass appears to be hopelessly too large to be in reach of the LHC, it fall squarely within 
the expected early reach of a hadron 100-TeV machine.    

The coupling $\hat{Y}^L_{12}$ -- not included in the fixed-point analysis -- is strongly constrained by the measurement
of the $\textrm{BR}(K^0_L\to \mu^+\mu^-)=(6.84\pm 0.11)\times 10^{-9}$\cite{PhysRevLett.84.1389}, 
which is practically saturated by the
absorptive long-distance contribution through $K_L^0\to \gamma \gamma$. One gets the 90\%~C.L. bound\cite{Mandal:2019gff}
\be
\left| \textrm{Re}(\hat{Y}^{L\ast}_{12}\hat{Y}^L_{22}) \right|\lesssim 1.2\times 10^{-5}\left(\frac{m_{S_3}}{\textrm{TeV}}\right)^2\,.\label{eq:KLMU}
\ee

In FP$_{2b}$, the scenario in best agreement with the experimental bounds, $\hat{Y}^L_{32}$ and
$\hat{Y}^L_{22}$ flow along irrelevant directions. The latter in particular becomes 
substantial at the low scale, being tied to the flow of the CKM matrix element $V_{33}$. 
If we assume that $\hat{Y}^L_{12}(M_{\textrm{Pl}})=0$ (independently of whether this happens along a relevant or irrelevant direction of flow), we get that
\be
\frac{d\hat{Y}^L_{12}}{dt}\simeq -\frac{y_t^2 V_{31}}{16\pi^2}\left(V_{32}\hat{Y}^L_{22}+V_{33}\hat{Y}^L_{32}\right)
\ee
predicts $\hat{Y}^L_{12}=1.6\times 10^{-6}$ at $Q=m_{S_3}$, which makes this scenario 
consistent with \refeq{eq:KLMU}.

Another potential low-energy bound comes from the experimental determination
of the $D_0\to \mu^+\mu^-$ branching ratio. The current measurement, $\textrm{BR}(D_0\to \mu^+\mu^-)<7.6\times 10^{-9}$ at the 95\%~C.L. at LHCb\cite{Aaij:2013cza} is already a few years old and might possibly 
be renewed with fresh data soon.
Roughly following, e.g., the analysis of Ref.\cite{Kowalska:2018ulj}, we can express the branching ratio in terms of the $S_3$ LQ parameters, assuming the $t$-channel exchange of the scalar $\phi_{1/3}$. 
One gets
\begin{equation}\label{eq:DMU}
\textrm{BR}(D_0\to \mu^+\mu^-)=
\tau_D\frac{f_D^2 M_D^5}{256\pi M_c^2}\left(\frac{2 M_{\mu} M_c}{M_D^2}\frac{\lam}{1-\lam^2/2}\,\frac{|\widetilde{Y}^L_{22}|^2}{2 m_{S_3}^2}\right)^2,
\end{equation}
where $\tau_D=4.1\times 10^{-13}\,\textrm{s}$ is the $D_0$ lifetime, $f_D=212\mev$ is the $D_0$ decay constant, 
$\lam=0.226$ is a Wolfenstein parameter, and $M_D$, $M_c$, $M_{\mu}$, are the $D$ meson, charm quark, and muon mass, respectively. Equation~(\ref{eq:DMU}) yields the $2\sigma$ bound
\be
|V_{22} \hat{Y}^L_{22}|\lesssim 0.63\, \frac{m_{S_3}}{\tev}\,,
\ee
which is currently not testing our scenarios at $Q=m_{S_3}\approx 5\tev$.

\section{Flavor anomalies in $\boldsymbol{b\to c}$ transitions}\label{sec:bcflav}

We move on to the second group of flavor anomalies receiving widespread attention in recent years: 
the deviations from the SM in the $R_{D^{(\ast)}}=\textrm{BR}(\bar{B}\to D^{(\ast)}\tau 
\nu)/\textrm{BR}(\bar{B}\to D^{(\ast)}l \nu)$ ratios, 
which have been observed at BELLE and LHCb\cite{Huschle:2015rga,Sato:2016svk,Hirose:2016wfn,Aaij:2015yra,Aaij:2017uff,Aaij:2017deq,Abdesselam:2019dgh,Belle:2019rba}, confirming previous hints from BaBar\cite{Lees:2012xj,Lees:2013uzd}. These  
anomalies in $b\to c$ transitions also imply a potential violation of lepton-flavor universality 
and admit an explanation
with LQs\cite{Sakaki:2013bfa,Fajfer:2015ycq,Dorsner:2016wpm,Becirevic:2016yqi,Li:2016vvp,Crivellin:2017zlb,Buttazzo:2017ixm,Angelescu:2018tyl,Marzocca:2018wcf,Iguro:2018vqb,Bansal:2018nwp,Yan:2019hpm,Popov:2019tyc,Cheung:2020sbq,Gherardi:2020det}.

While the NP potentially contributing to the $b\to s$ anomalies has to compete with SM loop effects, 
in the case of the charged-current $B$ anomalies 
that we discuss in this section the eventual presence of NP has to compete with 
the SM at the tree level. For equivalent Yukawa couplings, new states are thus naturally expected to be much lighter 
than in \refsec{sec:bsflav}, potentially in reach of the next round of LHC data. In this regard,
we do not attempt in this work to analyze NP models providing a simultaneous explanation to both the $b\to s$ and $b\to c$ anomalies, but rather keep \refsec{sec:bsflav} and \refsec{sec:bcflav} separated and independent
of one another. Moreover, we do not expect the trans-Planckian fixed-point analysis 
to provide in the case of $b\to c$ anomalies phenomenological information that is 
fundamentally enriching with respect to the well known findings of the numerous global fits existing in the literature\cite{Iguro:2018vqb,Murgui:2019czp,Bardhan:2019ljo,Blanke:2018yud,Blanke:2019qrx,Alok:2019uqc,Shi:2019gxi,Becirevic:2019tpx,Cheung:2020sbq}. 
We rather use this section to analyze the extent of the consistency of this NP with a gravity UV completion.

Several different operators in the weak effective theory are able to 
provide, alone and in combination, a $2\sigma$ explanation to the $b\to c$ anomalies, including 
the vector operator $\mathcal{O}_{V_1}=(\bar{c}\gamma^{\mu}P_L b)(\bar{\tau}\gamma_{\mu}P_L\nu)$, 
which is favored in single-operator scenarios and in combination with others.
$\mathcal{O}_{V_1}$ can be generated at the tree level by integrating out the scalar LQ $S_1$ or the vector LQ $U_1$. 
As we have mentioned in \refsec{sec:intro}, we focus on the scalar LQ to avoid having to introduce a further UV completion besides gravity.

The LQ $S_1$ is an SU(2) singlet. With respect to the $\textrm{SU(3)}_c\times\textrm{SU(2)}_L\times\textrm{U(1)}_Y$
gauge group its quantum numbers are $(\mathbf{\bar{3}},\mathbf{1},1/3)$.
One writes down the 
interaction Lagrangian in the SM quark mass basis in terms of left-chiral (unbarred) and right-chiral (barred) two-component spinors,
\be
\mathcal{L}\supset \left(\widetilde{Y}^L_{ij}u_{L,i} e_{L,j}-\hat{Y}^L_{ij}d_{L,i}\nu_{L,j}+Y^R_{ij}\bar{u}_{R,i}\bar{e}_{R,j}\right) S_1
+\textrm{H.c.}\,,
\ee
where repeated indices $i,j$ are summed over the SM generations,
and the Yukawa coupling matrices are again related to each other by the CKM matrix $V$, $\widetilde{Y}^L_{ij}=V^{\ast}_{ir}\hat{Y}^L_{rj}$.
Additional details of the $S_1$ model are given in Appendix~\ref{app:s1app}. 

When integrated out at its mass scale, $S_1$ generates the $C_{V_1}$ Wilson coefficient,
\be\label{eq:cv1}
C_{V_1}=\frac{\hat{Y}^L_{33}\widetilde{Y}^{L}_{23}}{4\sqrt{2}G_F V_{23}\, m_{S_1}^2}\,,
\ee
which features up- and down-like LQ Yukawa couplings (we limit ourselves to the case of real couplings). 
Global fits including the ratio $R(J/\psi)$ and the longitudinal polarization fraction of the $\tau$ lepton and $D^{\ast}$ meson in the data set\cite{Iguro:2018vqb,Cheung:2020sbq} point to the $2\sigma$ interval
\be
0.13\left(\frac{m_{S_1}^2}{\textrm{TeV}^2}\right)\lesssim \hat{Y}^L_{33}\widetilde{Y}^{L}_{23}\lesssim 0.36\left(\frac{m_{S_1}^2}{\textrm{TeV}^2}\right)\,.\label{c1bound}
\ee

We anticipate at this point that the fixed-point analysis predicts a product of left-chiral couplings 
$\hat{Y}^L_{33}\widetilde{Y}^{L}_{23}$ too small to fall within the interval~(\ref{c1bound}),
without finding at the same time $m_{S_1}$ in the region already excluded by the LHC. It bears resemblance in this
with the typical Yukawa values provided in Eqs.~(\ref{eq:FP1b})-(\ref{eq:FP3b}).
We must therefore seek for an alternative solution, also favored by the global fits,
characterized by a much smaller Wilson coefficient $C_{V_1}$, as long as it is accompanied by a substantial
Wilson coefficient
for the operator $\mathcal{O}_{S_2}=(\bar{c} P_R b)(\bar{\tau}P_L\nu)$,
$C_{S_2}\gsim 0.05$ (see, e.g., Ref.\cite{Cheung:2020sbq}).
In the $S_1$ LQ scenario we can generate 
\be\label{eq:cs2}
C_{S_2}=-\frac{\hat{Y}^L_{33} Y^{R}_{23}}{4\sqrt{2}G_F V_{23}\, m_{S_1}^2}\,,
\ee
which now involves the right-chiral couplings.

An additional complication arises from the fact that with substantial couplings of left and right chirality 
that connect the charm quark to the tau lepton we generate corrections to the mass of these two particles.
The Yukawa couplings of the charm quark and of the tau lepton must therefore be included in the fixed point analysis, 
and one ought to make sure that the corresponding masses are matched at the low energy.    
Note that the mass of the charm quark and the tau lepton are not very dissimilar 
from one another and therefore the corresponding Yukawa 
couplings do not feature a large hierarchy. This is fortunate, as it allows us to find, after running to the low scale, solutions 
that can match to the correct masses to a good approximation. 

\subsection{Fixed-point analysis\label{sec:FPS1}}

We employ the down-origin 
LQ Yukawa basis for the fixed-point analysis, as justified by the low-scale phenomenology. 
While the fit to the $b\to c$ anomalies, \refeq{c1bound}, 
does not provide explicit guidance in this regard, some other strong flavor constraints do. 
For example, the measurement of the branching ratio $\textrm{BR}(K^+\to \pi^+ \nu\bar{\nu})<1.85 \times 10^{-10}$
at the 90\% C.L. at NA62\cite{Trilov:2019yu,Mandal:2019gff}, implies
$\hat{Y}^L_{13}\hat{Y}^{L}_{23}\in\left[-3.7,8.3\right]\times10^{-4}$ $(m_{S_1}/\tev)^2$. 
In an up-type origin scenario for the couplings, this bound forces $\widetilde{Y}^L_{23}$ into a narrow interval, 
$(\widetilde{Y}^{L}_{23})^2 \approx \left[-4,2\right]\times 10^{-3}\,(m_{S_1}/\tev)^2$,
which ends up excluding potentially viable fixed-point solutions. 
Conversely, the points presented in the following paragraphs, obtained in the down-origin basis for the LQ Yukawa couplings, are not in tension with
flavor constraints.

The minimal set of couplings whose fixed point structure we are going to analyze consists of 12 independent parameters,
\be\label{model_par}
g_Y,\,g_2,\,g_3,\,y_t,\;y_b,\,y_c,\,y_\tau,\,\hat{Y}^L_{23},\,\hat{Y}^L_{33},\,Y^R_{23},\,Y^R_{33},\,V_{33}.
\ee
The RGEs for the $S_1$ plus SM system in the trans-Planckian regime are presented in Appendix~\ref{app:s1app}. Note that the element $Y^{R}_{33}$ of the right-handed LQ Yukawa matrix must be included in the analysis, as it is generated through the running even if its initial value is set to zero.  

In analogy with the fixed-point structure discussed in \refsec{sec:bsflav}, the non-abelian gauge couplings remain asymptotically free, while the abelian one develops a UV interactive fixed point, 
\be
g_3^\ast=0,\qquad g_2^\ast=0,\qquad g_Y^\ast=
\frac{12\pi}{5}\sqrt{\frac{2\,f_g}{5}}\,.
\ee
By fitting to the low-scale value of the hypercharge coupling one obtains $f_g=0.01$ and $g_Y^\ast=0.48$.

The situation in the Yukawa sector of the SM is, however, more involved. As the analysis 
includes the RGEs of the second and third generation of up-type quarks, we strive to preserve 
the hierarchy $y_t>y_c$ along the full RG flow to avoid the poles
in the beta function of the CKM matrix element $V_{33}$, cf.~\refeq{eq:ckms1} and Ref.\cite{Alkofer:2020vtb}. 
This implies that only three combinations of fixed-point values for top and charm Yukawa couplings are possible: 
$y_t^\ast=0$, $y_c^\ast=0$;  $y_t^\ast\neq0$, $y_c^\ast=0$; $y_t^\ast>y_c^\ast\neq 0$. 
The first of these cases does not yield solutions with LQ Yukawa couplings of the correct sign to fit the $b\to c$ anomalies. 
We thus focus on the cases with $y_t^\ast\neq 0$.

Another difference with the analysis of \refsec{sec:bsflav} is that we here need to fit four SM Yukawa couplings simultaneously. 
Like in \refsec{sec:bsflav}, matching $y_t$ and $y_b$ to their low-scale value determines the gravitational parameter $f_y$, so that we should demand
\be
y_c^\ast=0,\qquad y_\tau^\ast=0\,,
\ee
and look for a solution in which both associated directions are relevant. Additionally, $y_t$ and $y_c$ generate additive 
chiral symmetry-breaking 
contributions to the tau lepton Yukawa running, 
\be
\frac{dy_\tau}{dt} \approx \frac{-6}{16\pi^2}\left( y_c \hat{Y}^L_{23}Y^R_{23}+y_t \hat{Y}^L_{33} Y^R_{33}\right),
\ee
cf.~\refeq{eq:s1tau}. Therefore, in order to generate a UV fixed point for $y_{\tau}$ one of the elements $Y_{33}^{R}$, $\hat{Y}^L_{33}$ has to vanish at the fixed point. Since $Y_{33}^{R}$ does not enter directly 
the Wilson 
coefficients $C_{V_1}$ and $C_{S_2}$, we set $Y_{33}^{R\ast}=0$\,.

We are left with two possible combinations of fixed-point values in the SM Yukawa sector,
\bea
&\textrm{FP}_1:& y_t^\ast\neq 0, \quad y_b^\ast= 0, \quad  V_{33}^\ast= 0,\nonumber \\
&\textrm{FP}_2:& y_t^\ast\neq 0, \quad y_b^\ast\neq 0, \quad V_{33}^\ast= 0,\label{eq:FPS1SM}
\eea
accompanied by four different choices for the LQ Yukawa matrix elements,
\bea\label{eq:FPS1LQ}
&\textrm{FP}_a:& \hat{Y}_{23}^{L\ast}\neq 0, \quad \hat{Y}_{33}^{L\ast}= 0, \quad  Y_{23}^{R\ast}\neq 0,\nonumber \\
&\textrm{FP}_b:& \hat{Y}_{23}^{L\ast}\neq 0, \quad \hat{Y}_{33}^{L\ast}= 0, \quad  Y_{23}^{R\ast}= 0,\nonumber \\
&\textrm{FP}_c:& \hat{Y}_{23}^{L\ast}= 0, \quad \hat{Y}_{33}^{L\ast}\neq 0, \quad Y_{23}^{R\ast}\neq 0,\nonumber\\
&\textrm{FP}_d:& \hat{Y}_{23}^{L\ast}= 0, \quad \hat{Y}_{33}^{L\ast}\neq 0, \quad Y_{23}^{R\ast}= 0.
\eea
Note that we do not obtain any solutions corresponding to $\hat{Y}_{23}^{L\ast}\neq 0$ and $\hat{Y}_{33}^{L\ast}\neq 0$,
as they are not compatible with a relevant direction for $V_{33}$.

\begin{table*}[t]
\footnotesize
\begin{center}
\begin{tabular}{|c|c|ccccc|c|}
\hline
 & $f_y$ & $y_t^\ast$ & $y_b^\ast$ & $\hat{Y}^{L\ast}_{23}$ & $\hat{Y}^{L\ast}_{33}$ & $Y^{R\ast}_{23}$& Fit quality at $m_{S_1}$ \\
\hline
$1a$ & 0.0037 & $4\pi \frac{\sqrt{ 362 f_g + 1500 f_y}}{5 \sqrt{295}}$& 0 & $ 4\pi \frac{\sqrt{-249 f_g + 1250 f_y}}{5\sqrt{295}}$ & 0 &  $ 4\pi \frac{\sqrt{68 f_g + 65 f_y}}{\sqrt{295}}$  &  $2\,\sigma$ \\
$1b$ & 0.0017  & $4\pi \frac{\sqrt{14(27 f_g + 125 f_y)}}{5 \sqrt{355}}$& 0 & $ 4\pi \frac{\sqrt{219 f_g + 2000 f_y}}{5\sqrt{355}}$ & 0 &  0  & $\hat{Y}^L_{33}\hat{Y}^L_{23} < 0 $ \\
$1c$ & 0.0032 & $4\pi\frac{\sqrt{51 f_g +250 f_y}}{15\sqrt{5}}$ & 0 & 0 & $2\pi \frac{\sqrt{-66 f_g + 500 f_y}}{25}$ & $-4\pi \frac{\sqrt{141 f_g + 125 f_y}}{25}$ &  $\hat{Y}^L_{33}\hat{Y}^L_{23}$ too small\\
$1d$ & 0.0044 & $4\pi\frac{\sqrt{51 f_g +250 f_y}}{15\sqrt{5}}$ & 0 & 0 & $2\pi \frac{\sqrt{3 f_g + 25 f_y}}{5}$ & 0 &  $\hat{Y}^L_{33}\hat{Y}^L_{23}$ too small \\
\hline
\end{tabular}
\caption{Possible fixed points and the corresponding $f_y$ of the SM+$S_1$ system with the left- and right-handed couplings.  The low-energy prediction for the NP sector is shown in the last column on the right. }
\label{tab:S1FP}
\end{center}
\end{table*}

The fixed  points that are of phenomenological interest as possible solution to the $b\to c$ anomalies are summarized in \reftable{tab:S1FP}. We do not report in 
there fixed points of the type of FP$_2$ in \refeq{eq:FPS1SM}, 
as they predict too large a bottom mass at the low scale.

Fixed points of type FP$_{1c}$ and FP$_{1d}$ yield a low-scale set of solutions not dissimilar to FP$_{1b}$ 
of \refsec{sec:bsflav}. The dominant contribution to RG running takes in both cases the same form,
\be\label{eq:runY33}
\frac{d\hat{Y}_{33}^L}{dt}\simeq \frac{1}{16\pi^2}\hat{Y}_{23}^L\,y_t^2\,V_{33}\sqrt{1-V_{33}{}^2},
\ee
which leads to a reduction of the matrix element $\hat{Y}_{33}^L$ during its flow towards the IR. 
The low-energy value of $\hat{Y}^L_{33}$ is very small and results inconsistent 
with either \refeq{c1bound} or $C_{S_2}\gsim 0.05$ at the TeV scale (the low-scale value we obtain for $Y^R_{23}$ does not affect this conclusion). 

\begin{figure*}[t]
	\centering
		\includegraphics[width=0.45\textwidth]{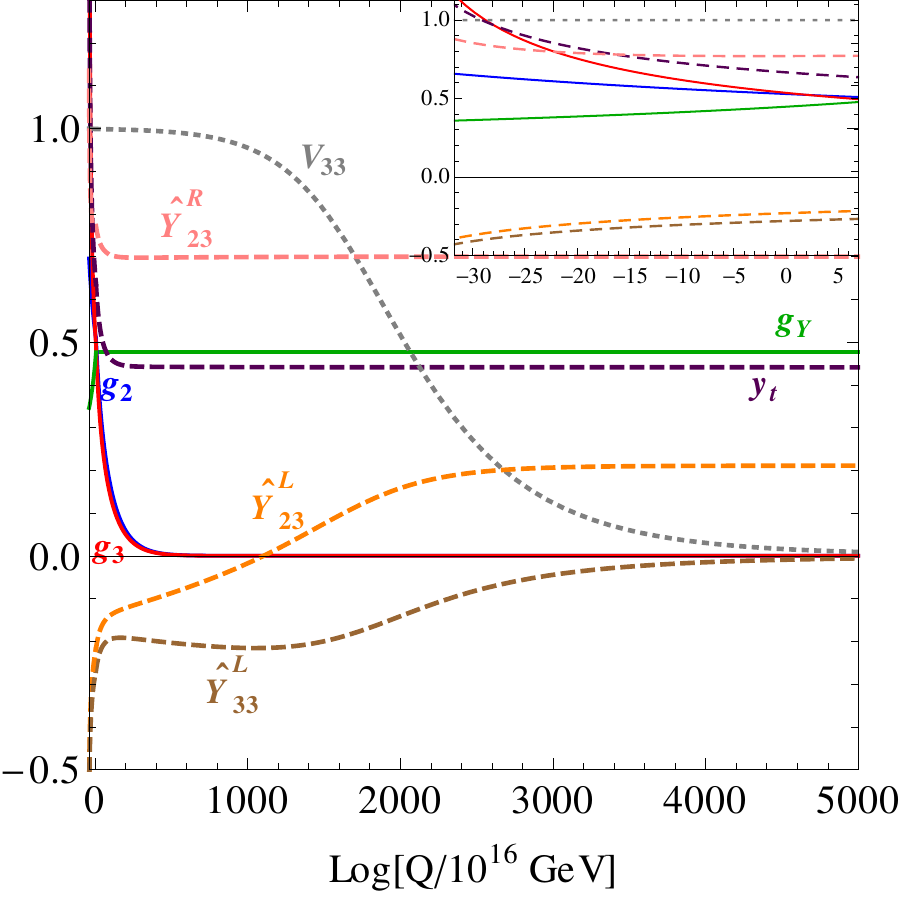}
	\caption{RG flow of the gauge and Yukawa couplings from the trans-Planckian energies down to the EWSB scale in the case of scenario $\textrm{FP}_{1a}$. The sub-Planckian flow is depicted in the inset panels.}
			\label{fig:run1c}
\end{figure*}

We now turn to discussing scenarios FP$_{1a}$ and FP$_{1b}$. The asymptotically free coupling $\hat{Y}^L_{33}$ is now associated with a relevant direction and its deviation from the UV fixed point is a free parameter of the theory. Since the main contribution to its running is given by \refeq{eq:runY33}, it is driven to negative values as soon as the CKM matrix element $V_{33}$ 
starts to depart from its fixed point. The RG flow of the coupling system from the vicinity of the UV fixed point towards the IR is shown for FP$_{1a}$ in \reffig{fig:run1c}. 

Fixed points FP$_{1a}$ and FP$_{1b}$ bear resemblance to FP$_{1a}$ of \refsec{sec:bsflav}, see also \reffig{fig:stream_S3a}(a). 
They do not, however, yield an identical low-scale phenomenology. The evolution of FP$_{1a}$ is
triggered mostly by the presence of the right-handed irrelevant coupling $Y^R_{23}$. 
Since its contribution to the running of $\hat{Y}^L_{23}$ is large and positive, 
\be
\frac{d\hat{Y}_{23}^L}{dt}\simeq \frac{1}{16\pi^2}\left(\hat{Y}_{23}^L\,(\hat{Y}_{23}^R)^2-2\hat{Y}_{33}^L\,y_t^2\,V_{33}\sqrt{1-V_{33}{}^2}\right),
\ee
see also 
\refeq{eq:s1yl23}, $\hat{Y}^L_{23}$ decreases towards the IR. As a consequence, we see that in FP$_{1a}$ 
both $\hat{Y}^L_{23}$ and $\hat{Y}^L_{33}$ can become negative and relatively sizable at the Planck scale. This behavior guarantees the partial consistency of the low-energy predictions of the model with the $b\to c$ flavor  
anomalies. Conversely, in scenario FP$_{1b}$  $\hat{Y}^L_{23}$ never becomes negative due to a small 
fixed-point value of the top Yukawa coupling. As a consequence, FP$_{1b}$ turns out to be not consistent with the low-energy phenomenology of the flavor anomalies. 

The Yukawa couplings $y_c$, $y_b$, $y_{\tau}$, which vanish at the fixed point, 
correspond to relevant directions in the coupling space, as long as $f_y$ exceeds the minimal value for which $y_b$ becomes relevant, as discussed in \refsec{sec:bsflav}. 
We checked that by fixing $f_y$ at this minimal value we predict the top mass at $Q=m_{S_1}=1.5\tev$ in very good agreement 
with the experimental measurement, $y_t(m_{S_1})=0.98$. However, the elements $\hat{Y}^{L}_{33}(m_{S_1})$, $\hat{Y}^{L}_{23}(m_{S_1})$, and $\hat{Y}^{R}_{23}(m_{S_1})$ are too small to fit the $b\to c$ anomalies. Therefore, we need to increase $f_y$ at the price of enhancing the top mass. 
Choosing $f_y=0.0036$, we obtain the following predictions for $\textrm{FP}_{1a}$:
\bea\label{eq:FPparam}
y_t(m_{S_1})=1.01, & & y_b(m_{S_1})=0.013,\nonumber\\
y_c(m_{S_1})=-0.009, & & y_{\tau}(m_{S_1})=0.01,\nonumber\\
\hat{Y}^L_{23}(m_{S_1})=-0.37, & & \hat{Y}^L_{33}(m_{S_1})=-0.41.\nonumber\\
Y^R_{23}(m_{S_1})=0.85\,. & & 
\eea

In all the scenarios discussed in this section $y_t$ corresponds to an irrelevant direction and the low-scale 
prediction for the top mass results by $\sim 10\%$ too large with respect to its experimentally measured value. We also observe that a fit to the tau lepton mass drives 
the charm mass to values by $\sim 40\%$ too large, even if the corresponding Yukawa couplings flow along relevant directions. This is an expected effect, due to the explicit breaking of chiral symmetry introduced 
by the presence of the $Y^R_{ij}$ couplings, which ties the running of $y_c$ to $y_{\tau}$. 
All these things considered, however, the fixed point of type FP$_{1a}$ features a satisfactory agreement  
between the quantum gravity UV completion and the full low-scale phenomenology. 

 \begin{figure*}[t]
	\centering
		\includegraphics[width=0.50\textwidth]{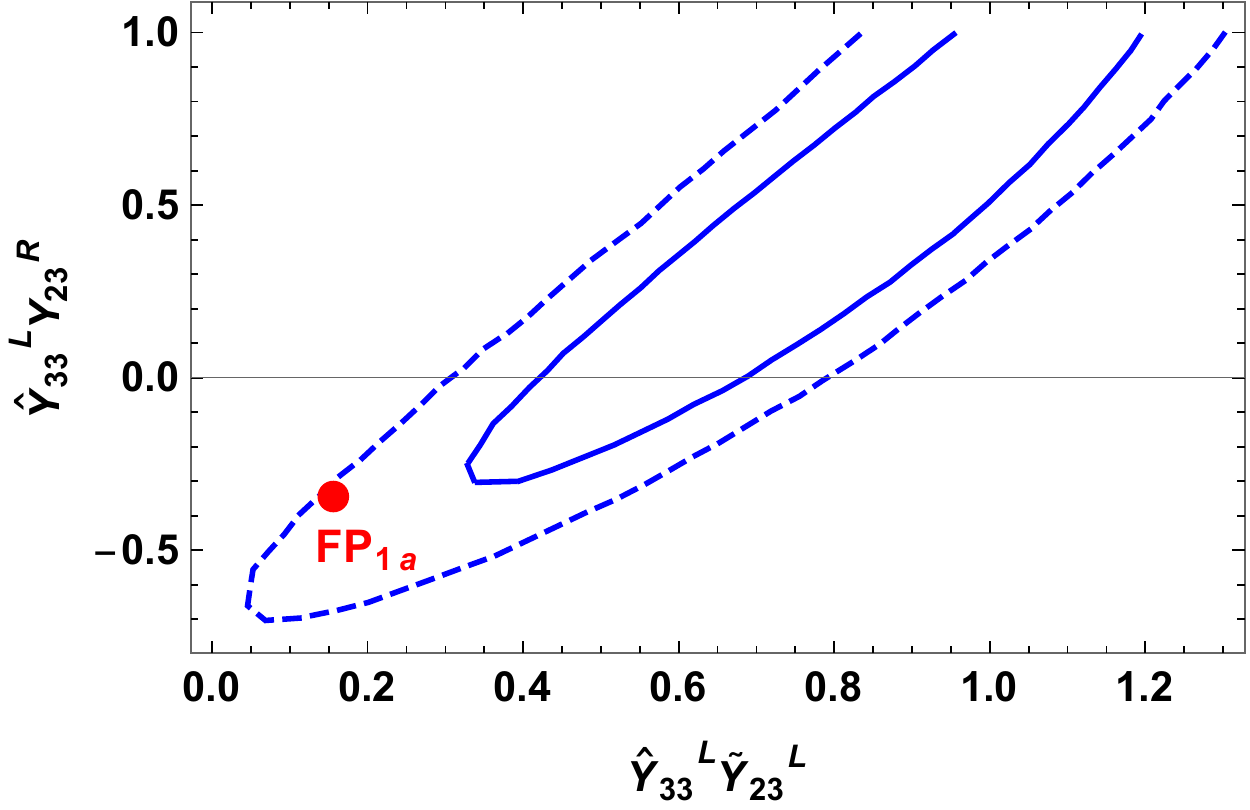}
	\caption{Solid (dashed) blue line gives the $1\sigma$ ($2\sigma$) confidence region of the $b\to c$ global fit of Ref.\cite{Cheung:2020sbq} in the plane
	($\hat{Y}^L_{33}\widetilde{Y}^L_{23}$, $\hat{Y}^L_{33} Y^R_{23}$) of the $S_1$ Yukawa couplings at $m_{S_1}=1.5\tev$. The red dot marks the position of FP$_{1a}$.}
			\label{fig:B2C_S1}
\end{figure*}

\subsection{Low-scale predictions}

We show in \reffig{fig:B2C_S1} 
the position of the fixed point FP$_{1a}$ in the plane ($\hat{Y}^L_{33}\widetilde{Y}^L_{23}$, $\hat{Y}^L_{33} Y^R_{23}$), 
compared to the $1\sigma$ (solid) and $2\sigma$ (dashed) regions of the global fit to $R_{D}/R_{D^{\ast}}$, $R(J/\psi)$, and the longitudinal polarization of the $\tau$ and $D^{\ast}$ presented in 
Ref.\cite{Cheung:2020sbq}. Like in Ref.\cite{Cheung:2020sbq}, the LQ mass is set at a reference value of $m_{S_1}=1.5\tev$.

Given the presence of relatively large left-handed LQ Yukawa couplings in \refeq{eq:FPparam}, 
the model is subject to a tight constraint from the measurement of the branching ratio 
$\textrm{BR}(B\to K^{(\ast)}\bar{\nu}\nu)$\cite{Sakaki:2013bfa}. The LQ contribution 
can be written as\cite{Buras:2014fpa}
\be
\mathcal{R}_{K}=\mathcal{R}_{K^{\ast}}=\frac{2}{3}+\frac{1}{3}\frac{\left|C_L^{\textrm{SM}}-2 \tilde{c}_{ql} \right|^2}{\left|C_L^{\textrm{SM}}\right|^2}\,,
\ee
where $C_L^{\textrm{SM}}=-6.38$\cite{Altmannshofer:2009ma} and
\be
\tilde{c}_{ql}\approx \frac{\left(5\tev\right)^2}{V_{tb}V_{ts}^{\ast}}\left(\frac{1}{4}
\frac{\hat{Y}^L_{23}\hat{Y}^L_{33}}{m_{S_1}^2}\right)\,.
\ee

By comparing the experimental determination, $\textrm{BR}(B\to K^{\ast 0}\,\bar{\nu}\nu)_{\textrm{exp}}<1.8\times 10^{-5}$ 
at the 90\%~C.L.\cite{Grygier:2017tzo}, with
$\mathcal{R}_{K^{\ast}}\cdot \textrm{BR}(B\to K^{\ast 0}\,\bar{\nu}\nu)_{\textrm{SM}}$,
where $\textrm{BR}(B\to K^{\ast 0}\,\bar{\nu}\nu)_{\textrm{SM}}=0.92\times 10^{-5}$, one gets
\be
-0.020 \leq\frac{\hat{Y}^L_{23}\hat{Y}^L_{33}}{m_{S_1}^2/\tev^2}\leq 0.061\,,
\ee
which places FP$_{1a}$ at the very edge of exclusion. Note that the model parameters imply that $\mathcal{R}_{K^{\ast}}=\mathcal{R}_{K}=2.4$ 
and that all decays of the type $\textrm{BR}(B\to K^{(\ast)}\bar{\nu}\nu)$ provide an equally good experimental venue 
to probe this kind of models.  

A potentially complementary signature for testing this scenario can be obtained by improving the 
precision of lepton universality measurements in the kaon sector. We define the universality as
\be\label{eq:LFUtau}
R^{K,\textrm{exp}}_{\tau/\mu}=\frac{\textrm{BR}(\tau^-\to K^-\nu)}{\textrm{BR}(K^-\to \mu^-\nu)}=0.01095\cdot(1\pm 0.015)\,,
\ee
as reported by the Particle Data Group (PDG)\cite{Tanabashi:2018oca}.
In the FP$_{1a}$ case we calculate
\be
R^{K}_{\tau/\mu}=\frac{\tau_{\tau}}{\tau_{K}}\frac{r_K^3}{2 r_{\mu}^2}\left(\frac{1-r_K^2}{r_K^2-r_{\mu}^2} \right)^2
\frac{\left(V_{12}+g_L\right)^2}{V_{12}^2}\,,
\ee
where $\tau_{\tau}=2.9\times 10^{-13}\,\textrm{s}$ and $\tau_{K}=1.25\times 10^{-8}\,\textrm{s}$ are the tau and kaon lifetimes, $V_{12}\approx \lam\approx 0.226$, $r_X=M_X/M_{\tau}$, and 
\be
g_L\equiv \frac{v_h^2}{m_{S_1}^2}\frac{\lam (\hat{Y}^L_{23})^2}{4}\,,
\ee
in terms of the Higgs vev, $v_h$. At $1\sigma$ we get
\be
|\hat{Y}^L_{23}|\leq 0.57\, \frac{m_{S_1}}{\tev}\,.
\ee
To test the value derived in \refeq{eq:FPparam}, $\hat{Y}^L_{23}=-0.37$ at $m_{S_1}=1.5\tev$, 
the measurement error in \refeq{eq:LFUtau} 
should be reduced to $\delta R^{K}_{\tau/\mu}/R^{K}_{\tau/\mu}\approx 2\times 10^{-3}$. 

Besides $R^{K}_{\tau/\mu}$, similar sensitivity to the $S_1$ model is expected in the observables $R^{D_s}_{\tau/\mu}$ and $B\to \tau \nu$, which provide additional testing ground for this scenario. 

Finally, one can use well known analytic approximations (see, e.g., Ref.\cite{Iguro:2018vqb}) to calculate $R_{D}$ and $R_{D^{\ast}}$ given the parameters of FP$_{1a}$ in \refeq{eq:FPparam}.  With $m_{S_1}=1.5\tev$ we obtain 
\be
R_{D}(\textrm{FP}_{1a})=0.341\,,\quad \quad R_{D^{\ast}}(\textrm{FP}_{1a})=0.270\,.
\ee
On the other hand, strictly speaking $m_{S_1}=1.5\tev$ is already excluded at the 95\%~C.L. by the most recent ATLAS data on LQ pair production in the gluon-gluon channel\cite{Aad:2020iuy}. The most recent bound, at $m_{S_1}>1.7\tev$, 
pushes FP$_{1a}$ slightly outside of the $2\sigma$ favored region for the $R_{D^{(\ast)}}$ anomalies, as the contours in \reffig{fig:B2C_S1} move a little to the right and down, whereas the benchmark point remains roughly in the same place.
Incidentally, at $m_{S_1}=1.7\tev$ the constraint from $\textrm{BR}(B\to K^{\ast 0}\,\bar{\nu}\nu)$ falls short 
of excluding the Yukawa values given in \refeq{eq:FPparam}. One recalculates  
\be
R_{D}(\textrm{FP}_{1a},\,1.7\tev)=0.332\,,\quad \quad R_{D^{\ast}}(\textrm{FP}_{1a},\,1.7\tev)=0.267\,.
\ee

At $m_{S_1}=1.7\tev$, the LHC is expected to reach in the $p\,p\to \mu^+\mu^-\textrm{jet}$ channel (quark-quark + quark-gluon production) a 95\%~C.L. sensitivity to $|Y^R_{22}|\approx 0.80$ with $300\,\textrm{fb}^{-1}$ of integrated luminosity\cite{Kowalska:2018ulj}. In the corresponding channel with final-state tau leptons -- appropriate to test 
the value given in \refeq{eq:FPparam}, $Y^R_{23}=0.85$ -- the LHC sensitivity drops, and is expected to exclude a Yukawa coupling larger by approximately 60\% than in the muon case\cite{Angelescu:2018tyl}. Still, a combination of the reach of gluon-gluon, quark-quark, quark-gluon production channels is likely to corner the scenario discussed here in the very near future.

\section{Summary and conclusions}\label{sec:summary}

In this paper we used the framework of asymptotically safe quantum gravity  
to derive predictions for the mass of scalar LQs as solutions to the experimental anomalies recorded in recent years 
in $b\to s $ and $b\to c$ transitions. 
Our predictions are obtained by embedding a SM extension with a single LQ in a trans-Planckian completion in which gravitational interactions induce corrections to the beta functions of the Lagrangian (gauge and Yukawa) couplings.  
The latter thus develop interactive fixed points in the extreme UV, which parametrize specific sets of boundary conditions at the Planck scale. The flavor phenomenology is then unambiguously determined by following the coupling flow to the EWSB scale. 

The low-scale phenomenological predictions follow from two main sources:
the presence of a fixed point in $g_Y$ and therefore the size and sign of $f_g\approx 0.01$; and the decoupling scale $M_{\textrm{Pl}}$. In this sense, the predictions described here can stem from genuine quantum gravity effects, even if the correction to the Yukawa couplings, $f_y$, has subdominant impact on the final result (additionally, $f_y$ cancels out altogether from the beta function of $V_{33}$). Being flavor-blind, gravitational interactions do not generate by their own very nature flavor signatures. 
They can, however, be a source for the Planck-scale boundary conditions   
of the Yukawa textures favored by the flavor phenomenology. Once the low-scale value of the NP Yukawa couplings is determined in this way, we found that by assuming a $2\sigma$ consistency with the neutral-current, $b\to s$ anomalies,
the predicted LQ mass lies in the range $4-7\tev$. 
These values are too large to be in reach of the high-luminosity LHC, but 
fall squarely within the early reach of a 100-TeV hadron collider, according to the most conservative estimates. 
Complementary signatures in flavor observables like $\textrm{BR}(K_L\to \mu^+\mu^-)$ or $\textrm{BR}(D_0\to \mu^+\mu^-)$ require significant increases in the experimental sensitivity with respect to the current bounds.   

The trans-Planckian fixed-point analysis encounters some additional complications when applied to LQ solutions to 
the charged-current, $b\to c$ anomalies, mostly due to the presence of explicit chiral-symmetry violating terms in the Yukawa coupling RGEs, 
which induce some tension in the low-energy fit to the fermion masses.    
The situation is, however, very optimistic from the observational point of view, as the LQ mass and Yukawa couplings are predicted to be at the very edge of the current LHC bounds, well within the reach of $300\,\textrm{fb}^{-1}$.  

The present study can be extended in several directions. 
First of all, the fixed-point analysis can be broadened to incorporate the impact of all three families of quarks and three real mixing angles of the CKM matrix. While we do not expect this extension to affect significantly our predictions on the low-energy NP phenomenology, it would be interesting to verify whether such a setup can still be consistent with the SM. 

In a similar spirit, one could include in the analysis the non-trivial flavor structure of the leptonic sector, which we neglected here for the sake of simplicity. The non-trivial form of the neutrino mixing matrix could potentially lead to lepton flavor-changing signatures, 
which will be tested in the next few years by experiments like MEG-II and others.
Following this direction of investigation, one would also need to address the mechanism for generating neutrino masses.

Finally, we did not incorporate in the fixed-point analysis the parameters characterizing the scalar potential of the model. We do not expect them to impact our findings, as the scalar couplings enter the RGEs of the gauge and Yukawa couplings at the third and the second loop, respectively, and in the perturbative domain their impact becomes highly loop-suppressed. 
However, an extended analysis of the scalar potential can be an interesting topic {\it per se}. 
The issue of its stability, as well as the correct prediction of the Higgs boson mass, could be analyzed and compared to the NP predictions and signatures presented here.  

LQ solutions to the flavor anomalies have been extensively studied in the literature from the point of view of their compatibility with a large number of constraints defined at about and below their mass scale. We find it encouraging that they can also show consistency with 
the theoretical framework of asymptotically safe quantum gravity, which is defined far above the scale that can be tested directly and was previously left unexplored in this context. 


\bigskip
\begin{center}
\textbf{ACKNOWLEDGMENTS}
\end{center}
\noindent 
We would like to thank Daniel Litim for helpful comments on the manuscript. 
KK is supported in part by the National Science Centre (Poland) under the research Grant No. 2017/26/E/ST2/00470.
EMS and YY are supported in part by the National Science Centre (Poland) under the research Grant No. 2017/26/D/ST2/00490. 
\bigskip

\appendix

\section*{Appendices}
\addcontentsline{toc}{section}{Appendices}
\section{Trans-Planckian renormalization of $\boldsymbol{S_3}$\label{app:s3app}}

We review in this appendix the notation we use for rotating the SM and LQ fields.
The LQ $S_3$ is characterized by the SM quantum numbers $(\mathbf{\bar{3}},\mathbf{3},1/3)$. 
One defines the $3\times 3$ Yukawa matrix $Y_L$ in the ``flavor'' (or gauge-symmetric) basis according to the Lagrangian
\be\label{eq:S3gilag}
\mathcal{L}\supset (Y_L)_{ij}\,Q^T_i (i\sigma_2) S_3 L_j+\textrm{H.c.}\,,
\ee
where $Q_i^T=(u_{L i},d_{L i})$ and $L_j=(\nu_{L j},e_{L j})^T$ are SU(2)$_L$ 
doublets of two-component left-chiral Weyl spinors, $S_3$ is the scalar LQ matrix
\be
S_3=\left( {\begin{array}{cc}
\phi_{1/3} & \sqrt{2}\phi_{4/3}\\
\sqrt{2}\phi_{-2/3} & -\phi_{1/3}
 \end{array} } \right),
\ee
and $\sigma_2$ is the second Pauli matrix. 
It is well known that the quantum numbers of $S_3$ allow for the presence of Lagrangian terms of the type $Q^T(i\sigma_2)S_3^{\dag}Q$, potentially leading to fast proton decay\cite{Arnold:2013cva}. We assume in this paper that these are forbidden by a symmetry (for example, conservation of baryon and/or lepton number). 

The LQ is characterized by a scalar potential connecting it to the SM Higgs doublet, $H$: 
\begin{equation}\label{eq:scalpot}
V(H,S_3)=\frac{1}{2}m_{S_3}^2\textrm{Tr} [S_3^{\dag}S_3]+\frac{1}{8}\lam_{S_3}\left(\textrm{Tr}[S_3^{\dag}S_3]\right)^2
+\frac{1}{2}\lam_{HS_3}\textrm{Tr} [S_3^{\dag}S_3]H^{\dag}H\,.
\end{equation}
The LQ fields do not develop a vacuum expectation value (vev).

We carry out the transformation from the flavor basis to the quark mass basis via the unitary rotation matrices $U_L$, $U_R$, $D_L$, and $D_R$. If $Y_U$, $Y_D$, and $Y_E$ are the Yukawa matrices of the SM in the flavor basis, the corresponding diagonal matrices $Y_u$ and $Y_d$ in the mass basis are given by   
\be
Y_u=U_L^{\dag} Y_U U_R\,,\quad Y_d=D_L^{\dag} Y_D D_R \,,
\ee
and the CKM matrix is defined as $V\equiv U^{\dag}_L D_L$.
We assume for simplicity that the charged lepton Yukawa matrix is trivially diagonal in the flavor basis so that 
$Y_e\equiv Y_E$. 

We work in the quark mass basis throughout this work. We further introduce several assumptions, which significantly 
simplify the analysis and yet produce interesting phenomenological signatures at the low scale.
\begin{itemize}
    \item We treat the Yukawa couplings of the SM and NP as real in flavor space
    \item Above the LQ mass scale, we work in the 2-quark family approximation. As the flavor anomalies only constrain the second and third quark family this is a reasonable approximation simplifying the fixed-point analysis. As a direct consequence, 
    the CKM matrix is orthogonal and described by one rotation angle for the purposes of the fixed-point analysis 
    \item We neglect the physics of neutrino masses and oscillations. In this approximation the SM charged lepton Yukawa matrices are rotated by the identity matrix
    \item We restrict the fixed-point analysis to the gauge-Yukawa system at one loop. We have checked numerically that the predictions 
    for the NP couplings change minimally under the addition of perturbative 2-loop contributions. 
    The parameters of the scalar potential do not enter at one loop in the gauge-Yukawa RGEs, and do not affect the phenomenological predictions presented in this work. One should keep in mind, however, that if the theory is expected to be complete the full fixed-point analysis should include the parameters of \refeq{eq:scalpot} and their coupling to the graviton, $f_{\lam}$. The full system should show consistency with the $125\gev$ Higgs mass and the stability of the scalar potential at all scales (see Refs.\cite{Bandyopadhyay:2016oif,Popov:2016fzr,Wetterich:2016uxm,Pawlowski:2018ixd} 
    for some related work). We leave the analysis of these important but separated issues for future work.   
\end{itemize}

One obtains the LQ Yukawa matrices introduced in \refsec{sec:bsflav} in the quark mass basis via unitary rotations
\be
\widetilde{Y}^L=U_L^T Y_L\,, \quad \hat{Y}^L=D_L^T Y_L\,,\label{S3flav}
\ee
which lead to $\widetilde{Y}^L=V^{\ast}\hat{Y}^L$. We emphasize that the matrices in \refeq{S3flav}
are not diagonal as we do not enforce any flavor symmetry.

Since the texture of the SM and NP Yukawa matrices is not fixed by additional flavor symmetries, 
their elements are all subject to RG-running modifications. Elements that are zero
in a particular basis at one scale, do not necessarily remain zero or small in the same basis 
at a different scale. One can however choose to remain at each and every scale in one specific basis, for example the  
quark mass basis that we select in this work, in which the SM Yukawa matrices are diagonal and the rotation matrices run accordingly. 
The explicit scale dependence of $U_L(t)$, 
$U_R(t)$, $D_L(t)$, and $D_R(t)$ will have to be factored into the RG flow.

Additionally, one can choose to carry out the fixed-point analysis in a preferential basis for the LQ Yukawa matrices: 
the \textit{up-origin} $\widetilde{Y}^L$, or the \textit{down-origin} $\hat{Y}^L$. Whether one or the other basis is chosen will result in one or another set of UV fixed point. Some of the fixed points will be consistent with the low-scale phenomenological constraints, whereas others will be in tension or outright excluded. In the case of the $b\to s$ anomalies, the nature of the constraint, \refeq{eq:bsbound}, leads to the natural choice of the down-origin basis for our analysis.
Following standard procedure (see, e.g., Ref.\cite{Alkofer:2020vtb}), 
we define the squared Yukawa matrices at the scale $t=\log Q$ in the flavor basis:
\be
M_U=Y_U Y_U^{\dag}\,,\quad \quad M_D=Y_D Y_D^{\dag}\,, \quad \quad M_L=Y_L Y_L^{\dag}\,.
\ee
The corresponding diagonal Yukawa matrices in the mass basis at the scale $t$ are given by
\be
Y_u^2(t)=U_L^{\dag}(t) M_U(t) U_L(t)\,, \quad \quad Y_d^2(t)=D_L^{\dag}(t) M_D(t) D_L(t) \,,
\ee
as well as the non-diagonal squared matrix
\be
\left(\hat{Y}^L \hat{Y}^{L\dag}\right)(t)=D_L^T(t) M_L(t)  D_L^{\ast}(t)\,.
\ee

We use \texttt{SARAHv4.12.2}\cite{Staub:2013tta} to derive the one-loop 
RG flow for the (squared) Yukawa matrices in the flavor basis and then use the 
unitarity of the rotation matrices to connect them to the RG flow in the mass basis. One writes
\bea
\partial_t Y_u^2+\left[Y_u^2,\left(\partial_t U_L^{\dag}\right) U_L\right]&=&U_L^{\dag}\left(\partial_t M_U\right)U_L\label{S3Yu}\\
\partial_t Y_d^2+\left[Y_d^2,\left(\partial_t D_L^{\dag}\right) D_L\right]&=&D_L^{\dag}\left(\partial_t M_D\right)D_L\label{S3Yd}\\
\partial_t  \left(\hat{Y}^L \hat{Y}^{L\dag}\right)+\left[ \hat{Y}^L \hat{Y}^{L\dag},\left(\partial_t D_L^T\right) D_L^{\ast}\right]&=&D_L^T\left(\partial_t M_L\right)D_L^{\ast}\,.\label{S3YL}
\eea
The l.h.s. of Eqs.~(\ref{S3Yu}) and (\ref{S3Yd}) is now recast 
as a sum of one diagonal and one purely off-diagonal matrix, the first of which features the SM Yukawa coupling beta function, whereas the second parametrizes the scale dependence of the rotation matrices. Note, on the other hand, that the l.h.s. of \refeq{S3YL} is not characterized by any specific texture, as both addends feature diagonal and off-diagonal elements.  

The scale dependence of Eqs.~(\ref{S3Yu}) and (\ref{S3Yd}) can be used to derive unambiguously
the flow of the absolute values of the CKM matrix elements\cite{Babu:1987im,Sasaki:1986jv,Barger:1992pk,Kielanowski:2008wm}.
We use the unitarity of the rotation matrices and the fact that $V=U_L^{\dag} D_L$ 
to write
\be
\frac{dV_{ij}}{dt}=\sum_{k=1,2,3}\left[\left(\partial_t U_L^{\dag}\, U_L\right)_{ik }V_{kj}-V_{ik}\left(\partial_t D_L^{\dag}\, D_L\right)_{kj}\right]\,,
\ee
and then use the identity 
\be
\frac{1}{|V_{ij}|^2}\frac{d|V_{ij}|^2}{dt}=2 \textrm{Re}\left(\frac{1}{V_{ij}}\frac{dV_{ij}}{dt}\right)
\ee
to be allowed to ignore safely the unknown imaginary part of the diagonal elements of the matrices 
$\partial_t U_L^{\dag}\, U_L$ and $\partial_t D_L^{\dag}\, D_L$.

We present here the one-loop gauge-Yukawa-CKM system of equations for $S_3$ in the down-origin basis and 2-family approximation. 
We only give the equations for the parameters affecting the low-scale phenomenology of the $b\to s$ anomalies. All other parameters can be 
considered to be zero and relevant at the UV fixed point.
\be
\frac{dg_3}{dt}=-\frac{13}{2}\frac{g_3^3}{16\pi^2}-f_g g_3
\ee
\be
\frac{dg_2}{dt}=-\frac{7}{6}\frac{g_2^3}{16\pi^2}-f_g g_2
\ee
\be
\frac{dg_Y}{dt}=\frac{43}{6}\frac{g_Y^3}{16\pi^2}-f_g g_Y
\ee
\begin{multline}\label{eq:yt_S3}
\frac{d y_t}{dt}=\frac{1}{16\pi^2}\left[3 y_b^2+\frac{9}{2} y_t^2-\frac{17}{12}g_Y^2-\frac{9}{4}g_2^2-8 g_3^2-\frac{3}{2}V_{33}{}^2 y_b^2\right.\\
\left.+\frac{3}{4}\left(V_{32}{}^2(\hat{Y}^L_{22})^2+2 V_{32}V_{33}\hat{Y}^L_{22}\hat{Y}^L_{32}
+V_{33}{}^2(\hat{Y}^L_{32})^2\right) \right]y_t-f_y y_t
\end{multline}
\begin{equation}
\frac{d y_b}{dt}=\frac{1}{16\pi^2}\left[\frac{9}{2} y_b^2+y_t^2-\frac{5}{12}g_Y^2-\frac{9}{4}g_2^2-8 g_3^2 -\frac{3}{2}V_{33}{}^2 y_t^2
+\frac{3}{4}(\hat{Y}^L_{32})^2 \right]y_b-f_y y_b
\end{equation}
\begin{multline}\label{eq:L22_S3}
    \frac{d\hat{Y}^L_{22}}{dt}=\frac{1}{16\pi^2}\left\{\left[\frac{7}{2}(\hat{Y}^L_{22})^2+ \frac{11}{4}(\hat{Y}^L_{32})^2-\frac{5}{6}g_Y^2-\frac{9}{2}g_2^2-4 g_3^2\right.\right.\\
\left.\left.+\frac{1}{2}y_t^2 V_{32}{}^2  \right]\hat{Y}^L_{22}+2 y_t^2 V_{32}V_{33}\hat{Y}^L_{32}\right\}-f_y \hat{Y}^L_{22}
\end{multline}
\begin{multline}\label{eq:L32_S3}
    \frac{d\hat{Y}^L_{32}}{dt}
    =\frac{1}{16\pi^2}\left\{\left[\frac{17}{4}(\hat{Y}^L_{22})^2+ \frac{7}{2}(\hat{Y}^L_{32})^2+\frac{1}{2}y_b^2-\frac{5}{6}g_Y^2-\frac{9}{2}g_2^2\right.\right.\\
\left.\left.-4 g_3^2 
    +\frac{1}{2}y_t^2 V_{33}{}^2\right]\hat{Y}^L_{32}-y_t^2 V_{32}V_{33}\hat{Y}^L_{22} \right\}-f_y \hat{Y}^L_{32}
\end{multline}
\begin{multline}\label{eq:V33_S3}
    \frac{d|V_{33}|}{dt}=\frac{V_{23}}{16\pi^2}
    \left[-\frac{3}{2}V_{23}V_{33}y_b^2
    +\frac{3}{4}\left(V_{22}V_{32}(\hat{Y}^L_{22})^2\right.\right.\\
    \left.\left.+V_{22}V_{33}\hat{Y}^L_{22}\hat{Y}^L_{32}+V_{23}V_{32}\hat{Y}^L_{22}\hat{Y}^L_{32}+V_{23}V_{33}(\hat{Y}^L_{32})^2 \right) \right]\\
-\frac{V_{32}}{16\pi^2}\left[\frac{3}{2}V_{32}V_{33}y_t^2
    -\frac{3}{4}\hat{Y}^L_{22}\hat{Y}^L_{32}\right]\,.
\end{multline}
Moreover, the orthogonality of the CKM matrix in the 2-family approximation leads to 
\be
V_{22}=V_{33}\,,\quad \quad V_{23}=-V_{32}=\sqrt{1-V_{33}{}^2}\,.
\ee

\section{Trans-Planckian renormalization of $\boldsymbol{S_1}$\label{app:s1app}}

We present here the RG flow of the $S_1$ gauge-Yukawa system in the trans-Planckian regime. 
$S_1$ carries the SM quantum numbers $(\mathbf{\bar{3}},\mathbf{1},1/3)$. 
In the flavor basis, the Lagrangian features NP Yukawa matrices of the \textit{left} (L) and \textit{right} (R) type. 
In terms of left-chiral (unbarred) and right-chiral (barred) Weyl fields one writes 
\be\label{eq:giS1}
\mathcal{L}\supset (Y_L)_{ij} Q_i^T(i\sigma_2)L_j S_1 +(Y_R)_{ij} \bar{u}_{Ri} \bar{e}_{R j} S_1+\textrm{H.c.}\,,
\ee
where a sum over repeated family indices $i,j$ is implied, 
$Q_i^T=(u_{L i},d_{L i})$ and $L_j=(\nu_{L j},e_{L j})^T$ are SU(2)$_L$ doublets 
and $\sigma_2$ is the second Pauli matrix. As was the case for $S_3$,
we assume that terms that are dangerous for proton decay are forbidden.

In dealing with \refeq{eq:giS1}, we adopt the assumptions introduced and justified in Appendix~\ref{app:s3app}. 
Namely, we restrict ourselves in the fixed-point analysis to real Yukawa couplings and the 2-family approximation; we neglect the physics of neutrinos and consider only diagonal charged-lepton matrices; we do not introduce the scalar potential sector parameters in the fixed-point
analysis.

The Yukawa matrices transform to the quark mass basis via the unitary rotations
\be
Y^R=U_R^T Y_R \quad \widetilde{Y}^L=U_L^T Y_L \quad \hat{Y}^L=D_L^T Y_L\,,\label{s1mass}
\ee
which yield $\widetilde{Y}^L=V^{\ast}\hat{Y}^L$.

As is explained in \refsec{sec:bcflav}, we perform the fixed-point analysis in the down-origin basis of the left-handed LQ Yukawa couplings, which is not in tension with the low-energy constraints. One must add to Eqs.~(\ref{S3Yu})-(\ref{S3YL}) the corresponding RGEs
\bea
\partial_t Y_u^2+\left[Y_u^2,\left(\partial_t U_R^{\dag}\right) U_R\right]&=&U_R^{\dag}\left(\partial_t M_{\overline{U}}\right)U_R\label{S1YuR}\\
\partial_t  \left(Y^R Y^{R\dag}\right)+\left[ Y^R Y^{R\dag},\left(\partial_t U_R^T\right) U_R^{\ast}\right]&=&U_R^T\left(\partial_t M_R\right)U_R^{\ast}\label{S1YR}\,,
\eea
obtained by rotating the squared matrices defined in the flavor basis,
\be
M_{\overline{U}}=Y_U^{\dag} Y_U\,, \quad \quad M_R=Y_R Y_R^{\dag}\,.
\ee

We present here the one-loop gauge-Yukawa-CKM system of equations for $S_1$ in the down-origin basis and 2-family approximation. 
We only give the equations for the parameters affecting the low-scale phenomenology of the $b\to c$ anomalies. All other parameters can be 
considered to be zero and relevant at the UV fixed point.
\be
\frac{dg_3}{dt}=-\frac{41}{6}\frac{g_3^3}{16\pi^2}-f_g g_3
\ee
\be
\frac{dg_2}{dt}=-\frac{19}{6}\frac{g_2^3}{16\pi^2}-f_g g_2
\ee
\be
\frac{dg_Y}{dt}=\frac{125}{18}\frac{g_Y^3}{16\pi^2}-f_g g_Y
\ee
\begin{multline}
\frac{d y_c}{dt}=\frac{1}{16\pi^2}\left\{\left[ 3 y_b^2+ \frac{9}{2} y_c^2 +3 y_t^2+ y_{\tau}^2-\frac{17}{12}g_Y^2-\frac{9}{4}g_2^2
-8g_3^2 \right.\right.\\
\left.\left.-\frac{3}{2}V_{23}{}^2 y_b^2+\frac{1}{2}(Y^R_{23})^2 +\frac{1}{2}\left(V_{22}{}^2(\hat{Y}^L_{23})^2+V_{23}{}^2(\hat{Y}^L_{33})^2\right.\right.\right.\\
\left.\left.\left.+2 V_{22}V_{23}\hat{Y}^L_{23}\hat{Y}^L_{33} \right) \right]y_c -2 y_{\tau} Y^R_{23}\left( V_{22}\hat{Y}^L_{23}+ V_{23}\hat{Y}^L_{33}\right)  \right\}-f_y y_c    
\end{multline}
\begin{multline}
\frac{d y_t}{dt}=\frac{1}{16\pi^2}\left\{\left[ 3 y_b^2+ 3 y_c^2 +\frac{9}{2} y_t^2+ y_{\tau}^2-\frac{17}{12}g_Y^2-\frac{9}{4}g_2^2-8g_3^2\right.\right.\\
\left.\left.-\frac{3}{2}V_{33}{}^2 y_b^2+\frac{1}{2}(Y^R_{33})^2+\frac{1}{2}\left(V_{32}{}^2(\hat{Y}^L_{23})^2+V_{33}{}^2(\hat{Y}^L_{33})^2\right.\right.\right.\\
\left.\left.\left.+2 V_{32}V_{33}\hat{Y}^L_{23}\hat{Y}^L_{33} \right) \right]y_t-2 y_{\tau}Y^R_{33}\left( V_{32}\hat{Y}^L_{23}+ V_{33}\hat{Y}^L_{33}\right) \right\}-f_y y_t      
\end{multline}
\begin{multline}
\frac{dy_b}{dt}=\frac{1}{16\pi^2}\left\{\left[\frac{9}{2} y_b^2 +3y_c^2 +3y_t^2 +y_{\tau}^2
-\frac{5}{12}g_Y^2-\frac{9}{4}g_2^2-8 g_3^2\right.\right.\\
\left.\left.-\frac{3}{2}\left(V_{23}{}^2 y_c^2+V_{33}{}^2 y_t^2\right) +\frac{1}{2}(\hat{Y}^L_{33})^2 \right]y_b \right\}-f_y y_b  
\end{multline}
\begin{multline}\label{eq:s1tau}
\frac{d y_{\tau}}{dt}=\frac{1}{16\pi^2}\left\{\left[3 y_b^2 +3y_c^2 +3y_t^2 +\frac{5}{2}y_{\tau}^2
-\frac{15}{4}g_Y^2-\frac{9}{4}g_2^2\right.\right.\\
\left.\left.+\frac{3}{2}\left((\hat{Y}^L_{23})^2+(\hat{Y}^L_{33})^2 \right) +\frac{3}{2}\left((Y^R_{23})^2+(Y^R_{33})^2\right)\right]y_{\tau}\right.\\
\left.-6\left[y_c Y^R_{23}\left(V_{22}\hat{Y}^L_{23}+V_{23}\hat{Y}^L_{33}\right)+y_t Y^R_{33}\left(V_{32}\hat{Y}^L_{23}+V_{33}\hat{Y}^L_{33}\right) \right]\right\}-f_y y_{\tau}
\end{multline}
\begin{multline}\label{eq:s1yl23}
\frac{d\hat{Y}^L_{23}}{dt}
=\frac{1}{16\pi^2}\left\{\left[4 (\hat{Y}^L_{23})^2+\frac{7}{2} (\hat{Y}^L_{33})^2+(Y^R_{23})^2+(Y^R_{33})^2-\frac{5}{6}g_Y^2\right.\right.\\
\left.\left.-\frac{9}{2}g_2^2-4 g_3^2
+\frac{1}{2}y_c^2V_{22}{}^2+\frac{1}{2}y_t^2 V_{32}{}^2 +\frac{1}{2}y_{\tau}^2  \right]\hat{Y}^L_{23}
+2\hat{Y}^L_{33}\left(y_c^2 V_{22}V_{23}\right.\right.\\
\left.\left.+y_t^2 V_{32}V_{33}\right)-2 y_{\tau} \left(y_c V_{22} Y^R_{23}+y_t V_{32} Y^R_{33} \right) \right\}-f_y \hat{Y}^L_{23}
\end{multline}
\begin{multline}
\frac{d\hat{Y}^L_{33}}{dt}
=\frac{1}{16\pi^2}\left\{\left[\frac{9}{2} (\hat{Y}^L_{23})^2+4 (\hat{Y}^L_{33})^2+(Y^R_{23})^2+(Y^R_{33})^2-\frac{5}{6}g_Y^2\right.\right.\\
\left.\left.-\frac{9}{2}g_2^2-4 g_3^2
+\frac{1}{2}y_b^2+\frac{1}{2}y_c^2V_{23}{}^2 +\frac{1}{2}y_t^2 V_{33}{}^2+\frac{1}{2}y_{\tau}^2  \right]\hat{Y}^L_{33}\right.\\
\left.-\hat{Y}^L_{23}\left(y_c^2 V_{22}V_{23}+y_t^2 V_{32}V_{33}\right)
-2 y_{\tau} \left(y_c V_{23} Y^R_{23}+y_t V_{33} Y^R_{33} \right) \right\}-f_y \hat{Y}^L_{33}
\end{multline}
\begin{multline}
\frac{dY^R_{23}}{dt}=\frac{1}{16\pi^2}\left\{\left[2 (\hat{Y}^L_{23})^2+2 (\hat{Y}^L_{33})^2+3 (Y^R_{23})^2+3 (Y^R_{33})^2\right.\right.\\
\left.\left.-\frac{1}{2}\frac{y_t^2+y_c^2}{y_t^2-y_c^2}(Y^R_{33})^2 -\frac{13}{3}g_Y^2-4 g_3^2
+y_c^2+y_{\tau}^2\right.\right.\\
\left.\left.+\frac{2y_t y_{\tau}}{y_t^2-y_c^2}(\hat{Y}^L_{23}V_{32}+\hat{Y}^L_{33}V_{33})Y^R_{33}  \right]Y^R_{23} -4 y_c y_{\tau}\left(V_{22}\hat{Y}^L_{23}+V_{23}\hat{Y}^L_{33} \right)\right.\\
\left.+\frac{3 y_c y_t}{y_t^2-y_c^2}y_b^2 V_{23}V_{33} Y^R_{33}+ \frac{2y_c y_{\tau}}{y_t^2-y_c^2}(Y^R_{33})^2(\hat{Y}^L_{23}V_{22}+\hat{Y}^L_{33}V_{23})\right.\\
\left.-\frac{y_c y_t}{y_t^2-y_c^2}\left[(\hat{Y}^L_{23})^2 V_{22}V_{32}+(\hat{Y}^L_{33})^2 V_{23}V_{33}\right.\right.\\
\left.\left.+\hat{Y}^L_{23}\hat{Y}^L_{33}(V_{23}V_{32}+V_{22}V_{33}) \right] Y^R_{33}  \right\}-f_y Y^R_{23}
\end{multline}
\begin{multline}
\frac{dY^R_{33}}{dt}=\frac{1}{16\pi^2}\left\{\left[2 (\hat{Y}^L_{23})^2+2 (\hat{Y}^L_{33})^2 +3 (Y^R_{23})^2+\frac{1}{2}\frac{y_t^2+y_c^2}{y_t^2-y_c^2}(Y^R_{23})^2\right.\right.\\
\left.\left.+3 (Y^R_{33})^2 -\frac{13}{3}g_Y^2-4 g_3^2
+y_c^2+y_{\tau}^2-\frac{2y_c y_{\tau}}{y_t^2-y_c^2}(\hat{Y}^L_{23}V_{22}+\hat{Y}^L_{33}V_{23})Y^R_{23}  \right]Y^R_{33}\right.\\
\left.-4 y_t y_{\tau}\left(V_{32}\hat{Y}^L_{23}+V_{33}\hat{Y}^L_{33} \right)-\frac{3 y_c y_t}{y_t^2-y_c^2}y_b^2 V_{23}V_{33} Y^R_{23}\right.\\
\left.- \frac{2y_t y_{\tau}}{y_t^2-y_c^2}(Y^R_{23})^2(\hat{Y}^L_{23}V_{32}+\hat{Y}^L_{33}V_{33}) +\frac{y_c y_t}{y_t^2-y_c^2}\left((\hat{Y}^L_{23})^2 V_{22}V_{32}\right.\right.\\
\left.\left.+(\hat{Y}^L_{33})^2 V_{23}V_{33}+\hat{Y}^L_{23}\hat{Y}^L_{33}(V_{23}V_{32}+V_{22}V_{33}) \right) Y^R_{23}  \right\}-f_y Y^R_{33}
\end{multline}
\begin{multline}\label{eq:ckms1}
\frac{d|V_{33}|}{dt}=\frac{V_{23}}{16\pi^2}\left[-\frac{3}{2}\frac{y_c^2+y_t^2}{y_t^2-y_c^2}
V_{33}V_{23}y_b^2+\frac{1}{2}\frac{2 y_t y_c}{y_t^2-y_c^2}Y^R_{33}Y^R_{23}\right.\\
\left.+\frac{1}{2}\frac{y_c^2+y_t^2}{y_t^2-y_c^2}\left(V_{22}V_{32}(\hat{Y}^L_{23})^2
    +V_{22}V_{33}\hat{Y}^L_{23}\hat{Y}^L_{33}+V_{23}V_{32}\hat{Y}^L_{23}\hat{Y}^L_{33}+V_{23}V_{33}(\hat{Y}^L_{33})^2 \right)
\right.\\
\left.-\frac{2 y_{\tau} y_t}{y_t^2-y_c^2}(\hat{Y}^L_{23}V_{22}+\hat{Y}^L_{33}V_{23})Y^R_{33}
-\frac{2 y_{\tau} y_c}{y_t^2-y_c^2}(\hat{Y}^L_{23}V_{32}+\hat{Y}^L_{33}V_{33})Y^R_{23}
\right]\\
-\frac{V_{32}}{16\pi^2}\left[\frac{3}{2}\left(y_t^2 V_{32}V_{33}+y_c^2 V_{22}V_{23}\right) -\frac{1}{2}\hat{Y}^L_{23}\hat{Y}^L_{33} \right]\,.
\end{multline}

As before, the orthogonality of the CKM matrix in the 2-family approximation leads to 
\be
V_{22}=V_{33}\,,\quad \quad V_{23}=-V_{32}=\sqrt{1-V_{33}{}^2}\,.
\ee

\bibliographystyle{utphysmcite}
\bibliography{mybib}

\end{document}